\documentclass[journal]{IEEEtran}
\usepackage{cite}
\usepackage{color}
\usepackage{algorithmic}
\usepackage{epstopdf}
\usepackage{textcomp}
\usepackage{enumerate}
\usepackage{amsfonts}
\usepackage{booktabs}
\usepackage{pifont}
\usepackage{enumerate}
\usepackage[cmex10]{amsmath}

\usepackage{mathabx}

\usepackage{soul}
\soulregister\cite7 
\soulregister\citep7 
\soulregister\citet7 
\soulregister\ref7 
\soulregister\eqref7 
\soulregister\pageref7 

\ifCLASSINFOpdf
   \usepackage[pdftex]{graphicx}
   \graphicspath{{../pdf/}{../jpeg/}}
   \DeclareGraphicsExtensions{.pdf,.jpeg,.png}
\else
   \usepackage[dvips]{graphicx}
   \graphicspath{{../eps/}}
   \DeclareGraphicsExtensions{.eps}
\fi
\begin{document}
%
\title{Joint Device Association, Resource Allocation and Computation Offloading in Ultra-Dense Multi-Device and Multi-Task IoT Networks}
%
%
%

\author{Tianqing~Zhou,~
        Yali~Yue,~
        Dong~Qin,~
        Xuefang~Nie,~
        Xuan~Li,~
        and Chunguo~Li~
\thanks{This work was supported by National Natural Science Foundation of China under Grant Nos. 61861017, 62171119, 61671144, 61861018, 61761030, 61961020, 61862025, 62001201, 61663010 and 61963017,  National Key Research and Development Program of China under Grant No. 2020YFB1807201, Foundation of Jiangxi Educational Committee of China under Grant Nos. GJJ180312 and GJJ180313.}
\thanks{T. Zhou, Y. Yue, X. Nie and X. Li are with the School of Information Engineering, East China Jiaotong University, Nanchang 330013, China (email: zhoutian930@163.com; yueyali1015@163.com;  Xuefangnie@163.com; lixuan0799@outlook.com).}
\thanks{D. Qin is with School of Information Engineering, Nanchang University, Nanchang 330031, China (e-mail: qindong@ncu.edu.cn).}
\thanks{C. Li is with School of Information Science and Engineering, Southeast University, Nanjing 210096, China (email: chunguoli@seu.edu.cn)}
}

\maketitle

\begin{abstract}
With the emergence of more and more applications of Internet-of-Things (IoT) mobile devices (IMDs), a contradiction between mobile energy demand and limited battery capacity becomes increasingly prominent. In addition, in ultra-dense IoT networks, the ultra-densely deployed small base stations (SBSs) will consume a large amount of energy. To reduce the network-wide energy consumption and extend the standby time of IMDs and SBSs, under the proportional computation resource allocation and devices' latency constraints, we jointly perform the device association, computation offloading and resource allocation to minimize the network-wide energy consumption for ultra-dense multi-device and multi-task IoT networks. To further balance the network loads and fully utilize the computation resources, we take account of multi-step computation offloading. Considering that the finally formulated problem is in a nonlinear and mixed-integer form, we utilize the hierarchical adaptive search (HAS) algorithm to find its solution. Then, we give the convergence, computation complexity and parallel implementation analyses for such an algorithm. By comparing with other algorithms, we can easily find that such an algorithm can greatly reduce the network-wide energy consumption under devices' latency constraints.
\end{abstract}

\begin{IEEEkeywords}
MEC, IoT networks, multi-device, multi-task, resource allocation, device association, computation offloading.
\end{IEEEkeywords}

%
\IEEEpeerreviewmaketitle

\section{Introduction}
%
%
%
%
\IEEEPARstart{W}{i}th the development of information technologies and Internet of Things (IoT), more and more new applications have emerged, e.g., pilotless automobiles, virtual reality, augmented reality, panoramic video \cite{HGuo2019Oct,JZhao2020Aug}. It is easy to find that most of these emerging applications are computing-intensive and delay-sensitive. However, in the reality, IoT mobile devices (IMDs) cannot meet the computation power or storage requirements of these applications well. To tackle this problem, the external computation or storage capacity is introduced. Consequently, the cloud computing framework was proposed, which provides some services for IMDs (users) by collecting much computation and storage resources into a data center \cite{TZhao2020May,JZhao2019Aug}.
\par
As we know, the cloud computing framework is highly centralized. In such a framework, the computation data of IMDs needs to be updated to a cloud center at first, and then the computation tasks are executed in this center. However, a large amount of data transfer will incur severe network congestion and high operation cost. In order to shorten the distance between IMDs and computation center, fog computing was presented \cite{MChiang2017Apr}, which is closer to the data source than cloud computing. In fog computing networks, IMDs' computation tasks and applications can be tackled at the edge of networks, rather than on the cloud. Following this, to further promote the concept of ``local processing power" of fog computing, mobile edge computing (MEC) emerged. MEC greatly shortens the distance between IMDs and computation center, and execute IMDs' computation tasks and applications in lower latency and energy consumption, and higher security \cite{TXTran2017Apr}.
\par
To further shorten the distance between IMDs and computation center, ultra-dense small base stations (SBSs) are deployed into heterogeneous cellular networks (HCNs) \cite{JZhao2019Oct,TZhou2018Mar,TZhou2017Oct}, where all SBSs and macro BSs (MBSs) are implemented with edge computing servers. Although such a framework can greatly enhance the service coverage, and balance the computation tasks among BSs, the deployment of ultra-dense SBSs will result in abundant energy consumption \cite{TZhou2018,YDai2018Dec,SYu2021Feb,TZou2021Oct}.
\par
Evidently, under the limited network resources, it is a hot topic of how to reduce the network-wide energy consumption and extend the standby time of mobile terminal devices (IMDs) and SBSs, which jointly considers the computation offloading, user (device) association and resource allocation. Significantly, the computation offloading problem may be to decide whether IMDs' computation tasks are executed on edge servers or mobile terminals, or it may be to decide how much IMDs' computation tasks are executed on edge servers and mobile terminals. In addition, the device association problem is to decide which BS should be selected by one IMD.

\subsection{Related Work}
Although the spectrum sharing can improve the spectrum utilization efficiency in 5G wireless heterogeneous networks, it results in severe interference among users. To effectively reduce the network interference and enhance the system performance, a variety of resource management methods and strategies have been advocated. Under the latency constraints of users, Tan \textit{et al.} in \cite{TanL2021} introduced orthogonal frequency division multiple access (OFDMA) into MEC networks, and jointly performed the task offloading and resource allocation to minimize the total energy consumption of all users. After introducing non-orthogonal multiple access technologies (NOMA) into MEC networks, Xue \textit{et al.} in \cite{XueJ2021} jointly considered the task offloading and resource allocation to maximize the task processing capacity for multi-server and multi-task MEC networks.
\par
Meanwhile, there exist a lot of efforts made on the network scenarios with multi-user and multi-server. Ding \textit{et al.} in \cite{DingY2020Jul} developed a code-oriented partitioning offloading mechanism to decide the CPU frequency, execution location, and transmission power of users while minimizing the weighted sum of computation time and energy consumption for multiple users and multiple MEC servers under limited computation power and waiting task queue. At last, the formulated problem was converted into a convex form, and a decentralized computation offloading algorithm was established. Cheng \textit{et al.} in \cite{ChengQQ2020} jointly investigated the computation offloading and resource allocation to minimize users' energy consumption and task delay for uplink ultra-dense NOMA-MEC-enabled networks. Then, a novel mean field-deep deterministic policy gradient (MF-DDPG) algorithm was proposed to achieve the optimal solution of formulated problem.
\par
It is easy to find that the aforementioned researches concentrate on the multi-user systems, but not the multi-task ones. With the constantly updating of applications, many users need to process multiple tasks. Evidently, it is necessary for us to study the algorithm design for multi-user multi-task systems.  Chen M \textit{et al.} in \cite{ChenM2018Oct} jointly optimized the offloading decision, the computational and communicational resource allocation to minimize the overall cost of computation, energy and delay of users for a general multi-user MCC (mobile cloud computing) system. In such a system, each user has multiple independent computation tasks. At last, an efficient three-step algorithm was established to find the locally optimal solution of formulated problem. Ning \textit{et al.} in \cite{NingZ2019Jun} studied a problem of minimizing total execution delay of computation tasks of all users in the scenarios with multiple dependent tasks, utilized the branch and bound algorithm to solve single-user computation offloading problem, and designed an iterative heuristic algorithm to solve multi-user one. Chen W \textit{et al.} in \cite{ChenW2019Sep} investigated a multi-user multi-task computation offloading problem of maximizing overall revenue for MEC networks with energy harvesting, used the Lyapunov optimization to determine the energy harvesting policy, and then introduced centralized and distributed greedy maximum scheduling algorithms to resolve the formulated problem. Chen J \textit{et al.} in \cite{ChenJ2021} mainly solved the dependent task offloading problem of minimizing average energy-time cost in multi-user scenarios, which was modeled as a Markov decision process. Then, an Actor-Critic mechanism was proposed to attain the solution of formulated problem. In addition, Guo S \textit{et al.} in \cite{GuoS2019Feb} studied a problem of minimizing energy-efficiency cost under the task-dependency requirements and completion time deadline constraints, and presented a distributed dynamic offloading and resource scheduling (eDors) algorithm to obtain its solution. Guo H \textit{et al.} in \cite{HGuo2019Oct} jointly optimized the offloading decision, local computation capacity and power allocation to minimize the weighted time and energy consumption for ultra-dense IoT networks with multi-task, multi-device and multi-server implementation.
\par
Furthermore, with the development of emerging industries, the demand for information is gradually increasing. In many places, e.g., shopping malls and stations, the task requests are highly intensive. At this time, MEC servers may have insufficient computation resources for served users, which will cause long user delay and high energy consumption. To solve this kind of problem, more and more scholars concentrated on joint computation offloading and resource allocation. Kan \textit{et al.} in \cite{KanT2018Jun} considered the allocation of both radio resources and computation resources to minimize the network cost, and proposed a heuristic algorithm. Labidi \textit{et al.} in \cite{LabidiW2015Jun} jointly performed the resource scheduling and computation offloading to minimize the average energy consumption of users, and proposed offline dynamic programming approaches. Hu \textit{et al.} in \cite{HuS2020Feb} considered a problem of joint request offloading and resource scheduling to minimize the response delay of requests for MEC-enabled ultra-dense networks, and designed the cross-direction genetic algorithm (GA) to achieve the solution of formulated problem. Elgendy \textit{et al.} in \cite{ElgendyA2020Dec} tried to minimize the weighted sum of energy consumption of users by jointly optimizing the task compression, offloading and security for multi-user multi-task MEC networks, utilized linearization and relaxation approaches to achieve the solution of formulated problem. Dai \textit{et al.} in \cite{YDai2018Dec} jointly performed the computation capacity optimization, user association, power control and computation offloading to minimize the network-wide energy consumption for multi-user multi-task networks.
\par
Among the efforts mentioned above, a few of them take account of computation offloading in ultra-dense IoT networks, and very few of them consider the two-step computation offloading. In addition, very few efforts have been made towards the energy efficiency optimization under the proportional computation resource allocation.
\subsection{Contributions and Organization}
In this paper, a problem of minimizing network-wide energy consumption is formulated, which jointly optimizes the device association, power control, computation offloading and resource partition. Specifically, the main contributions of this paper can be listed as follows.
\begin{enumerate}[1)]
\item \textit{Multi-Step Computation Offloading in Ultra-Dense IoT Networks:} We consider the both one-step and two-step computation offloading in ultra-dense IoT Networks. In the one-step computation offloading, IMDs can be associated with MBS and offload computation tasks to MBS directly. In the two-step computation offloading, IMDs firstly need to offload partial tasks to SBSs and then SBSs may offload partial tasks to MBS. To enhance the performance of multi-step computation offloading, we try to optimize the frequency (band) partition used for avoiding the inter-tier interference. In addition, the equal frequency utilization is advocated to eliminate intra-tier and intra-cell interference. Evidently, the multi-step computation offloading in such an interference management mechanism should be a new investigation.
\item \textit{Computation Offloading Problem Formulation under Proportional Computation Resource Allocation:} Under the proportional computation resource allocation, a problem of network-wide energy consumption minimization is formulated for a multi-device multi-task system. In such a problem, the device association, power control, computation offloading and frequency band resource partition are jointly optimized under the IMDs' latency constraints. It is evident that this is a new formulation and investigation.
\item \textit{Solution of Formulated Problem:} Considering that the formulated problem is in a nonlinear and mixed-integer form, we utilize the hierarchical adaptive search (HAS) algorithm to find its solution. To this end, we need to make some proper changes on the gene encoding, selection, crossover and mutation of individuals. In addition, unlike the existing efforts, the optimized parameters are encoded into two types of chromosomes, which have different lengths.
\item \textit{The Convergence and Computation Complexity Analyses:} As for the algorithms developed in this paper, we make some detailed investigations on the convergence, computation complexity and parallel implementation analyses.
\end{enumerate}
\par
The rest of this paper is organized as follows. Section \ref{sec 2} introduces system model including network model, communication model, computation model and multi-task model; Section \ref{sec 3} gives the formulation of a problem of minimizing network-wide energy consumption under IMDs' latency constraints; Section \ref{sec 4} utilizes HAS to solve the formulated problem; Section \ref{sec 5} provides the detailed algorithm analysis including convergence, complexity and parallel implementation analyses; Section \ref{sec 6} gives the simulation results and analyses; Section \ref{sec 7} provides the conclusion and further discussion.
\section{System Model}\label{sec 2}
In this section, we firstly show the network model, i.e., ultra-dense heterogeneous cellular networks with multiple IMDs, tasks and MEC servers. Then, the communication, computation and multi-task models are described in detail.
\subsection{Network Model}
\begin{figure}[!t]
\centering
\centerline{\includegraphics[width=3.8in]{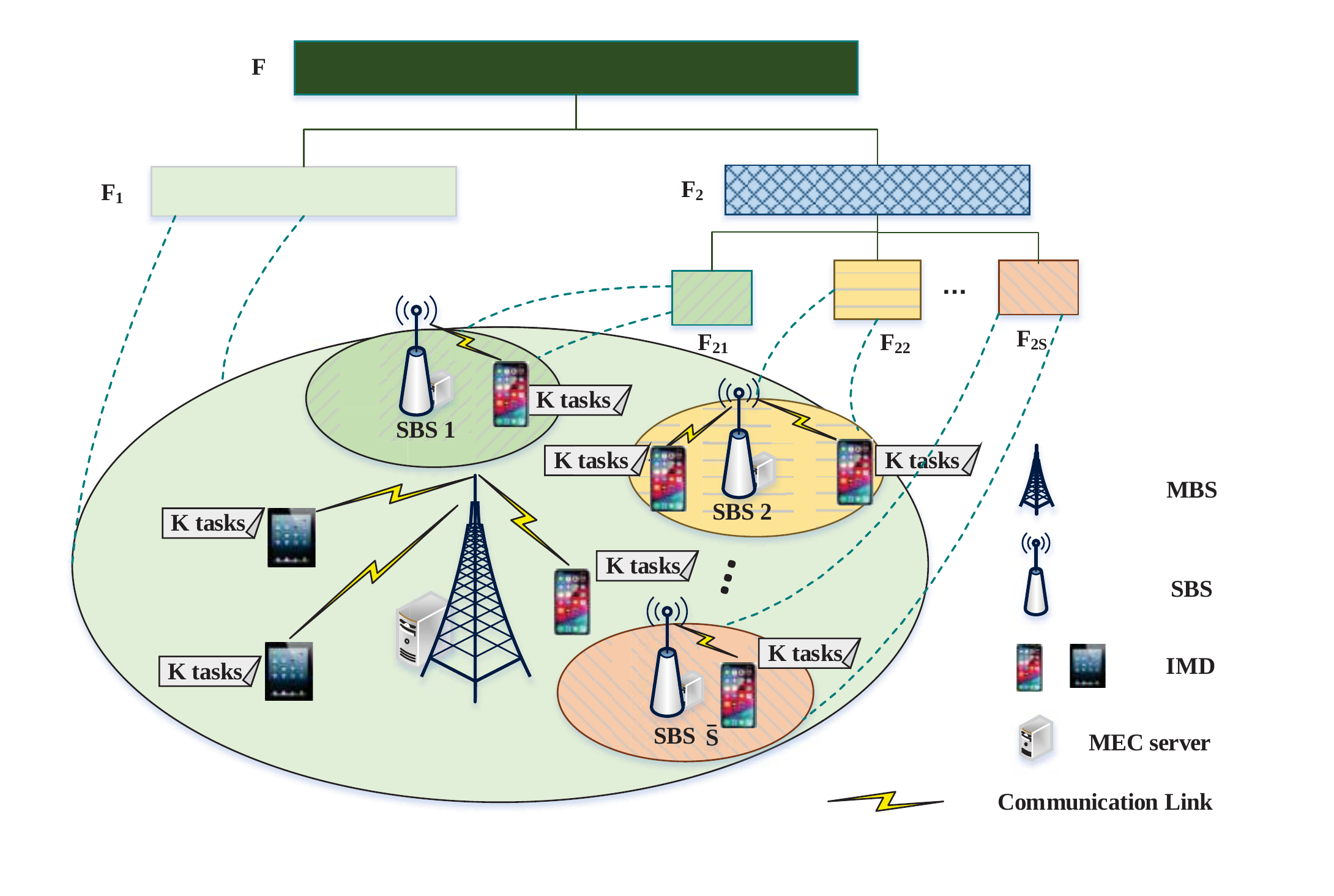}}
\caption{MEC-enabled ultra-dense IoT Networks with multi-task, multi-device and multi-server implementation.}
\label{fig1}
\end{figure}

In this paper, we concentrate on MEC-enabled ultra-dense IoT networks with multi-task, multi-device and multi-server implementation, which is illustrated in Fig.\ref{fig1}. In such networks, ultra-dense SBSs are deployed into each macrocell, and any BS is equipped with one MEC server. Generally, in any macrocell, the number of SBSs is greater than or equal to the one of IMDs. Without loss of generality, we consider one MBS and $\bar{S}$ SBSs in MEC-enabled ultra-dense networks, where the set of SBSs is represented as $\bar{\mathcal{S}}=\big\{ 1,2,\cdots ,\bar{ S} \big\}$; the index of MBS is given by $0$; $\mathcal{S}=\bar{\mathcal{S}}\cup \{0\} $ indicates the set of all BSs; $U$ IMDs lie in the set  $\mathcal{U}= \left\{1,2, \cdots ,U\right\}$. In this paper, we assume that all SBSs are connected to the MBS via wired links, and any IMD has a computing-intensive and latency-sensitive application to be executed within a certain deadline. In addition, we consider that each application of any IMD has $K$ relatively independent tasks, which are denoted by $\mathcal{K}=\left\{ 1,2,\cdots ,K \right\}$.
\par
As illustrated in Fig.\ref{fig1}, an effective interference management mechanism is introduced to eliminate the network interference in this paper. Specifically, the whole frequency band ${F}$ is cut into two parts including ${{F}_{1}}$ and ${{F}_{2}}$, which are used for MBSs and SBSs respectively. Then, the frequency band ${{F}_{2}}$ is allocated to SBSs equally, and the frequency band of each SBS is allocated to its associated IMDs equally. Significantly, the widths of frequency bands ${F}$, ${{F}_{1}}$ and ${{F}_{2}}$ are $\mathcal W $, $\lambda \mathcal W$ and $\left( 1\text{-}\lambda  \right)\mathcal W$ respectively, where $0\le \lambda \le 1$ is the frequency band partitioning factor. In this way, the inter-tier interference can be cancelled, the intra-tier interference can be eliminated, and the intra-cell interference can be avoided completely. Although the spectrum utilization ratio of such an interference management mechanism is relatively low, the frequency band allocated to IMDs should be sufficient since each SBS often serves at most one IMD in ultra-dense IoT Networks.
\subsection{Communication Model}
Since the data size of computation results downloaded from any BS is very small, the time used for downloading them can be negligible. That is to say, we just need to concentrate on the uplink transmission.
\par
If IMD $i$ is associated with SBS $j$, the uplink signal-to-interference-plus-noise ratio (SINR) from IMD $i$ to SBS $j$ can be given by
\begin{equation}\label{eq1}
{{\gamma }_{ij}}=\frac{{{p}_{i}}{{\hbar}_{ij}}}{{{\sigma }^{2}}},\forall j\in \bar{\mathcal{S}},
\end{equation}
where ${{p}_{i}}$ represents the transmission power of IMD $i$; ${{\hbar}_{ij}}$ denotes the channel gain between IMD $i$ and SBS $j$; ${{\sigma }^{2}}$ is the noise power.
\par
Under the interference management mechanism mentioned above, the uplink data rate ${{R}_{ij}}$ from IMD $i$ to SBS $j$ can be given by
\begin{equation}\label{eq2}
{{R}_{ij}}=\frac{\left( 1-\lambda  \right)\mathcal W}{\bar{S}\sum\nolimits_{m\in \mathcal{U}}{{{x}_{mj}}}}\log_2\left( 1+{{\gamma }_{ij}} \right),
\end{equation}
where $\sum\nolimits_{m\in \mathcal{U}}{{{x}_{mj}}} $ represents the number of IMDs associated with SBS $j$; ${\left( 1-\lambda  \right)\mathcal W}/\left({\bar{S}\sum\nolimits_{m\in \mathcal{U}}{{{x}_{mj}}}}\right)$ represents the bandwidth of any IMD associated with SBS $j$; ${{x}_{ij}}$ denotes the association index between IMD $i$ and SBS $j$; ${{x}_{ij}}=1 $ if IMD $i$ is associated with SBS $j$, ${{x}_{ij}}=0$ otherwise.
\par
If IMD $i$ is associated with MBS, the uplink signal-to-interference-plus-noise ratio (SINR) from IMD $i$ to MBS can be given by
\begin{equation}\label{eq3}
{{\gamma }_{i0}}=\frac{{{p}_{i}}{{\hbar}_{i0}}}{{{\sigma }^{2}}}
\end{equation}
where ${{\hbar}_{i0}}$ represents the channel gain between IMD $ i $ and MBS.
\par
Similarly, the uplink data rate ${{R}_{i0}}$ from IMD $i$ to MBS can be given by
\begin{equation}\label{eq4}
{{R}_{i0}}=\frac{\lambda \mathcal W}{\sum\nolimits_{m\in \mathcal{U}}{{{x}_{m0}}}}\log_2(1+{{\gamma }_{i0}}),
\end{equation}
where $\sum\nolimits_{m\in \mathcal{U}}{{{x}_{m0}}}$ represents the number of IMDs associated with MBS; ${\lambda \mathcal W}/\left({\sum\nolimits_{m\in \mathcal{U}}{{{x}_{m0}}}}\right)$ represents the bandwidth of any IMD associated with MBS.
\subsection{Computation Model }
We assume that any IMD $i$ has an application composed of $K$ tasks. In addition, the $k$-th task in this application can be expressed as ${\mathcal{ D}_{ik}}\overset{\Delta }{\mathop{=}}\,\left( {{d}_{ik}},{{c}_{ik}} \right)$, where ${{d}_{ik}}$ denotes the data size of the $ k$-th computation task of IMD $i$; ${{c}_{ik}}$ is the number of CPU cycles used for computing one bit of task ${\mathcal{D}_{ik}}$. In addition, the execution time of any IMD $i$ cannot exceed its deadline $T_{i}^{\max }$.
\par
In this paper, we consider that the computation offloading procedure can include the following two steps. In the first step, the part of $k$-th task of IMD $ i $ is offloaded to SBS $j$ for processing. In the second step, the part of $k$-th task offloaded to SBS $j$ is offloaded to MBS for processing. Specifically, when IMD $i$ is associated with SBS $j$, ${{\bar{d}}_{ijk}}$ is offloaded to this BS for processing and ${{d}_{ik}}-{{\bar{d}}_{ijk}}$ is calculated locally. Then, $ {{\hat{d}}_{ijk}} $ is offloaded to MBS for processing, and the remaining ${{\bar{d}}_{ijk}}-{{\hat{d}}_{ijk}}$ is processed at SBS $j$. Certainly, we can let IMD $i$ be associated with MBS. At this time, ${{\bar{d}}_{i0k}}$ is offloaded to MBS for processing and ${{d}_{ik}}-{{\bar{d}}_{i0k}}$ is calculated locally.
\par
Next, the time and energy consumption for computation offloading will be discussed in different scenarios.
\par
\textit{\textbf{1) Local computation:}} When IMD $i$ is associated with BS $j$, the amount of $k$-th task processed locally is ${{d}_{ik}}-{{\bar{d}}_{ijk}}$, and the local execution time  $T_{_{ijk}}^{LOC}$ used for processing $k$-th task of IMD $ i $ associated with BS $j$ can be given by
\begin{equation}\label{eq5}
T_{_{ijk}}^{LOC}=\frac{\left( {{d}_{ik}}-{{{\bar{d}}}_{ijk}} \right){{c}_{ik}}}{{{f}_{ik}}},
\end{equation}
where ${{f}_{ik}}$ represents the computation capacity allocated by IMD $ i $ to the $k$-th task.
\par
For ease of algorithm design, the computation capacity of IMD $i$ is allocated to $k$-th task according to the CPU occupation ratios of all tasks. Specifically, the computation capacity ${{f}_{ik}}$ of IMD $i$ is allocated to $k$-th task can be given by
\begin{equation}\label{eq6}
{{f}_{ik}}=\frac{\sum\nolimits_{j\in \mathcal{S}}{{{x}_{ij}}\left( {{d}_{ik}}-{{{\bar{d}}}_{ijk}} \right){{c}_{ik}}}}{\sum\nolimits_{j\in \mathcal{S}}{\sum\nolimits_{l\in \mathcal{K}}{{{x}_{ij}}\left( {{d}_{il}}-{{{\bar{d}}}_{ijl}} \right){{c}_{il}}}}}F_{i}^{UE},
\end{equation}
where $ F_{i}^{UE} $ represents the total computation capacity of IMD $i$.
\par
When IMD $i$ is associated with BS $j$, the local computation energy consumption $E_{ijk}^{LOC}$ used for executing the remaining amount of $k$-th task can be given by
\begin{equation}\label{eq7}
	E_{ijk}^{LOC}=\bar{\varepsilon } \left( {{d}_{ik}}-{{{\bar{d}}}_{ijk}} \right){{c}_{ik}}f_{ik}^{2},
\end{equation}
where $\bar{\varepsilon }$ is the effective switched capacitance depending on the chip architecture.
\par
\textit{\textbf{2) Offloading to SBS}}: When IMD $i$ adopts two-step computation offloading to complete its $k$-th task, the time used for this type of operation includes four parts. In detail, the first part is the uplink transmission time used for uploading tasks to SBSs, the second one is the execution time of tasks at SBS, the third one is the uplink transmission time used for uploading tasks to MBS, and the last one is the execution time of tasks at MBS. Therefore, under the two-step computation offloading, when IMD $ i $ is associated with SBS $j$, the time $ T_{ijk}^{SBS} $ used for completing its $k$-th task can be given by
\begin{equation}\label{eq8}
T_{ijk}^{SBS}=\frac{{{{\bar{d}}}_{ijk}}}{{{R}_{ij}}}+\frac{\big( {{{\bar{d}}}_{ijk}}-{{{\hat{d}}}_{ijk}} \big){{c}_{ik}}}{{{{\bar{f}}}_{ijk}}}+\frac{{{{\hat{d}}}_{ijk}}}{{{r}_{0}}}+\frac{{{{\hat{d}}}_{ijk}}{{c}_{ik}}}{{{{\bar{f}}}_{i0k}}},
\end{equation}
where ${{r}_{0}}$ denotes the wired backhauling rate from SBS to MBS; ${{\bar{f}}_{ijk}}$ is the computation capacity allocated to $k$-th task of IMD $i$ by SBS $j$, and $ {{\bar{f}}_{i0k}} $ is the one allocated to $k$-th task of IMD $i$ by MBS; in the right side of \eqref{eq8}, the first, second, third and fourth terms are the uplink transmission time used for uploading tasks to SBSs, the execution time of tasks at SBS, the uplink transmission time used for uploading tasks to MBS, and the execution time of tasks at MBS respectively.
\par
When IMD $i$ is associated with SBS $j$, the computation capacity of SBS $j$ is allocated to $k$-th task of IMD $i$ according to the CPU occupation ratios of all tasks. Specifically, under the two-step computation offloading, when IMD $ i $ is associated with SBS $j$, the computation capacity ${{\bar{f}}_{ijk}}$ of SBS $j$ is allocated to $k$-th task of IMD $i$ can be given by
\begin{equation}\label{eq9}
	{{\bar{f}}_{ijk}}=\frac{\big( {{{\bar{d}}}_{ijk}}-{{{\hat{d}}}_{ijk}} \big){{c}_{ik}}}{\sum\nolimits_{m\in \mathcal{U}}{{{x}_{mj}}}\sum\nolimits_{l\in \mathcal{K}}{\big( {{{\bar{d}}}_{mjl}}-{{{\hat{d}}}_{mjl}} \big){{c}_{ml}}}}F_{j}^{BS},
\end{equation}
where $ F_{j}^{BS}  $ represents the total computation capacity of SBS $j$.
\par
Since SBSs can upload tasks to MBS for processing if IMDs are associated with these SBSs, and IMDs can also directly upload them to MBS for execution if these IMDs are associated with MBS, the data processed at MBS should include the following two parts. When some IMDs are associated with SBSs, the first part refers to the amount of data uploaded by these associated SBS, and it is given by $ {{A}_{1}}=\sum\nolimits_{m\in \mathcal{U}}{\sum\nolimits_{j\in \bar{\mathcal{S}}}{{{x}_{mj}}\sum\nolimits_{l\in \mathcal{K}}{{{{\hat{d}}}_{mjl}}{{c}_{ml}}}}} $. When some IMDs are associated with MBS, the second part refers to the amount of data uploaded by IMDs associated with MBS, and it is given by $ { A }_{2}=\sum\nolimits_{m\in \mathcal{U}}{{{x}_{m0}}\sum\nolimits_{l\in \mathcal{K}}{{\bar{d}_{m0l}}{{c}_{ml}}}} $. Consequently, according to the CPU occupation ratios of all tasks, the computation capacity $ {{\bar{f}}_{i0k}} $ of MBS is allocated to $k$-th task of IMD $i$ can be given by
\begin{equation}\label{eq10}
{{\bar{f}}_{i0k}}=\frac{\sum\nolimits_{j\in\bar{\mathcal{S}}}{{{x}_{ij}}{{{\hat{d}}}_{ijk}}{{c}_{ik}}}+{{x}_{i0}}{\bar{d}_{i0k}}{{c}_{ik}}}{{{A}_{1}}+{{A}_{2}}}F_{0}^{BS},
\end{equation}
where $ F_{0}^{BS} $ is the total computation capacity of MBS.
\par
When IMD $i$ adopts two-step computation offloading to complete its $k$-th task, the energy consumption used for this type of operation should include the following four parts. Specifically, the first part is the energy consumption used for uploading tasks to SBSs, the second one is the execution energy consumption of tasks at SBS, the third one is the energy consumption used for uploading tasks to MBS, and the last one is the execution energy consumption of tasks at MBS. Therefore, under the two-step computation offloading, when IMD $ i $ is associated with SBS $j$, the energy consumption $E_{ijk}^{SBS}$ used for executing its $k$-th task can be given by
\begin{equation}\label{eq11}
E_{ijk}^{SBS}=\frac{{{p}_{i}}{{{\bar{d}}}_{ijk}}}{{{R}_{ij}}}+\big( {{{\bar{d}}}_{ijk}}-{{{\hat{d}}}_{ijk}} \big){{c}_{ik}}{\hat{\varepsilon }_{j}}+\frac{\tilde{\varepsilon } {{{\hat{d}}}_{ijk}}}{{{r}_{0}}}+{{\hat{d}}_{ijk}}{{c}_{ik}}{\hat{\varepsilon }_{0}},
\end{equation}
where ${\hat{\varepsilon }_{j}}$ and ${\hat{\varepsilon }_{0}}$ represent the energy consumption of each CPU cycle at SBS and MBS respectively; $\tilde{\varepsilon }$ denotes the power consumption per second on wired line; in the right side of \eqref{eq11}, the first, second, third and fourth terms are the energy consumption used for uploading tasks to SBSs, the execution energy consumption of tasks at SBS, the energy consumption used for uploading tasks to MBS, and the execution energy consumption of tasks at MBS respectively.
\par
\emph{\textbf{3) Offloading to MBS:}} When IMD $i$ adopts one-step computation offloading to complete its $k$-th task, it is associated with MBS. At this time, the time $ T_{i0k}^{MBS} $ used for this type of operation can be given by
\begin{equation}\label{eq12}
T_{i0k}^{MBS}=\frac{{{{\bar{d}}}_{i0k}}}{{{R}_{i0}}}+\frac{{{{\bar{d}}}_{i0k}}{{c}_{ik}}}{{{{\bar{f}}}_{i0k}}},
\end{equation}
where the first term in the right side of \eqref{eq12} is the uplink transmission time used for uploading tasks to MBS, and the second one is the execution time of tasks at MBS.
\par
When IMD $i$ adopts one-step computation offloading to complete its $k$-th task, the energy consumption $E_{i0k}^{MBS}$ used for this type of operation can be given by
\begin{equation}\label{eq13}
	E_{i0k}^{MBS}=\frac{{{p}_{i}}{{{\bar{d}}}_{i0k}}}{{{R}_{i0}}}+{{\bar{d}}_{i0k}}{{c}_{ik}}{\hat{\varepsilon }_{0}},
\end{equation}
where the first term in the right side of \eqref{eq13} is the energy consumption used for uploading tasks to MBS, and the second one is the execution energy consumption of tasks at MBS.
\subsection{ Multi-task Model}
To meet the practical implementation, we assume that all computation tasks are executed sequentially. That is to say, for any IMD $i$, its $k$-th task can be executed only when its first $k-1$ tasks are completed. In addition, we assume that the local execution and computation offloading are performed simultaneously. Therefore, the total time $T_{i}^{Sq}$ of IMD $i$ used for completing its computation tasks is the maximum of local execution and computation offloading time, and it can be given by
\begin{equation}\label{eq14}
T_{i}^{Sq}=\sum\limits_{k\in \mathcal{K}}{\max \bigg( \sum\limits_{j\in \mathcal{S}}{{{x}_{ij}}T_{_{ijk}}^{LOC}},\sum\limits_{j\in \bar{\mathcal{S}}}{{{x}_{ij}}T_{_{ijk}}^{SBS}}+{{x}_{i0}}T_{_{i0k}}^{MBS} \bigg)}.
\end{equation}
However, the total energy consumption $E_{i}^{Sq}$ of IMD $i$ used for completing its computation tasks is the sum of local execution and computation offloading energy consumption, and it can be given by
\begin{equation}\label{eq15}
E_{i}^{Sq}=\sum\limits_{k\in \mathcal{K}}{\bigg( \sum\limits_{j\in \mathcal{S}}{{{x}_{ij}}E_{_{ijk}}^{LOC}}\text{+}\sum\limits_{j\in \bar{\mathcal{S}}}{{{x}_{ij}}E_{_{ijk}}^{SBS}}+{{x}_{i0}}E_{_{i0k}}^{MBS} \bigg)}.
\end{equation}
\section{PROBLEM FORMULATION}\label{sec 3}
In order to reduce the network-wide energy consumption, and extend the standby time of mobile terminal devices (IMDs) and SBSs, we jointly perform the device association, computation offloading and resource allocation to minimize the network-wide energy consumption under IMD' latency constraints for ultra-dense multi-device and multi-task IoT Networks. It is noteworthy that the proportional computation resource allocation is utilized before the problem formulation. Specifically, the optimization problem is formulated as
\begin{equation}\label{eq16}
	\begin{split}
		& \underset{\mathbf{X},\mathbf{p},\mathbf{\bar{D}},\mathbf{\hat{D}},\lambda }{\mathop{\min}}\,  E\left( \mathbf{X},\mathbf{p},\mathbf{\bar{D}},\mathbf{\hat{D}},\lambda  \right)=\sum\nolimits_{i\in \mathcal{U}}{E_{i}^{Sq}} \\
		&\text{s.t. }{{C}_{1}}:T_{i}^{Sq}\le T_{i}^{\max },\forall i\in \mathcal{U}, \\
		& {{C}_{2}}:\sum\nolimits_{j\in \mathcal{S}}{{{x}_{ij}}}=1,\forall i\in \mathcal{U}, \\
		& C{}_{3}:\theta\le {{p}_{i}}\le p_{i}^{\max },\forall i\in \mathcal{U}, \\
		& {{C}_{4}}:{{x}_{ij}}\in \left\{ 0,1 \right\},\forall i\in \mathcal{U},j\in \mathcal{S}, \\
		& {{C}_{5}}:\theta\le \sum\nolimits_{j\in \bar{\mathcal{S}}}{{{x}_{ij}}{{{\hat{d}}}_{ijk}}}\le \sum\nolimits_{j\in \bar{\mathcal{S}}}{{{x}_{ij}}{{{\bar{d}}}_{ijk}}}\le {{d}_{ik}},\forall i\in \mathcal{U},k\in \mathcal{K}, \\
		& {{C}_{6}}:\theta\le {{x}_{i0}}{{{\bar{d}}}_{i0k}}\le {{d}_{ik}},\forall i\in \mathcal{U},k\in \mathcal{K}, \\
		& {{C}_{7}}:\theta\le \lambda \le 1, \\
	\end{split}
\end{equation}
where $\mathbf{X}=\{ {{x}_{ij}},\forall i\in \mathcal{U},\forall j\in \mathcal{S} \}$, $\mathbf{p}=\{ {p_{i}},\forall i\in \mathcal{U} \}$, $\mathbf{\bar{D}}=\{ {\bar{d}_{ijk}},\forall i\in \mathcal{U},\forall j\in \mathcal{S},\forall k\in \mathcal{K} \}$ and $\mathbf{\hat{D}}=\{ {\hat{d}_{ijk}},\forall i\in \mathcal{U},\forall j\in \mathcal{S},\forall k\in \mathcal{K} \}$; $\theta$ takes a small enough value to avoid the zero division, e.g., ${{10}^{-20}}$; ${{C}_{1}}$ indicates that the task execution time of IMD $i$ cannot be greater than the deadline $ T_{i}^{\max} $; ${{C}_{2}}$ and ${{C}_{4}}$ show that any IMD can just be associated with one BS; ${{C}_{3}}$ gives the lower bound ($\theta$) and upper bound ($ p_{i}^{\max } $) of transmission power of IMD $i$; ${{C}_{7}}$ gives the lower bound ($\theta$) and upper bound (1) of frequency band partitioning factor. In addition, as shown in ${{C}_{5}} $, when IMD $i$ is associated with SBS $j$, this IMD can offload $ {{{\bar{d}}}_{ijk}}$ bit of $k$-th task to SBS $j$, and then SBS $j$ can offload ${{{\hat{d}}}_{ijk}}$ bit of its received partial $k$-th task to MBS. Evidently, $ {{{\bar{d}}}_{ijk}}$ and ${{{\hat{d}}}_{ijk}}$ should be greater than or equal to $\theta$, but less than or equal to the data size ${{d}_{ik}}$ of $k$-th task of IMD $i$. Meanwhile, ${{{\hat{d}}}_{ijk}}$ should be less than or equal to $ {{{\bar{d}}}_{ijk}}$. As revealed in ${{C}_{6}}$, when IMD $i$ is associated with MBS, this IMD can offload $ {{{\bar{d}}}_{i0k}}$ bit of $k$-th task to MBS. It is evident that $ {{{\bar{d}}}_{i0k}}$ should be greater than or equal to $\theta$, but less than or equal to the data size ${{d}_{ik}}$ of $k$-th task of IMD $i$.
\par
It is easy to find that the formulated problem \eqref{eq16} is in a nonlinear and mixed-integer form, and the optimized parameters are also highly coupling. That means such a problem is in a nonconvex form. In ultra-dense IoT Networks, the problem \eqref{eq16} is often a large-scale mixed-integer nonlinear programming one. At this time, it is evident that an exhaustive searching method utilized for testing all possible solutions should be impractical and infeasible.
\section{ALGORITHM DESIGN}\label{sec 4}
To solve the problem \eqref{eq16}, we utilize the hierarchical adaptive search (HAS) algorithm \cite{TZou2021Oct} to find its solution, which is the combination of adaptive genetic algorithm (GA) with diversity guided mutation (DGM) \cite{EtterD1982} and adaptive particle swarm algorithm (PSO) algorithm \cite{ShiY1998}. In the whole algorithm, adaptive GA with GDM is firstly used for coarse-grained search, and adaptive PSO is then employed for fine-grained search. Compared with the efforts in \cite{FGuo2018Dec}, such an algorithm can avoid the premature convergence and improve the convergence speed, and finally achieve a better solution.
\subsection{Adaptive GA with DGM}
It is easy to find that GA is essentially a series of operations on the chromosome patterns. In other words, the selection operation is used to inherit good patterns from the current population to the next generation, the crossover operation is utilized to reorganize patterns, and the mutation operation is adopted to mutate the patterns \cite{EtterD1982}. Through these genetic operations, the chromosome pattern evolves toward a better direction gradually, and the optimal solution of formulated problem can be obtained finally. That is to say, in order to solve the problem \eqref{eq16}, GA starts with a set of initial feasible solutions, and then repeatedly performs the selection, crossover and mutation operations until an acceptable solution of problem achieves or GA converges.
\begin{figure}[!t]
\centering
\centerline{\includegraphics[width=3.8in]{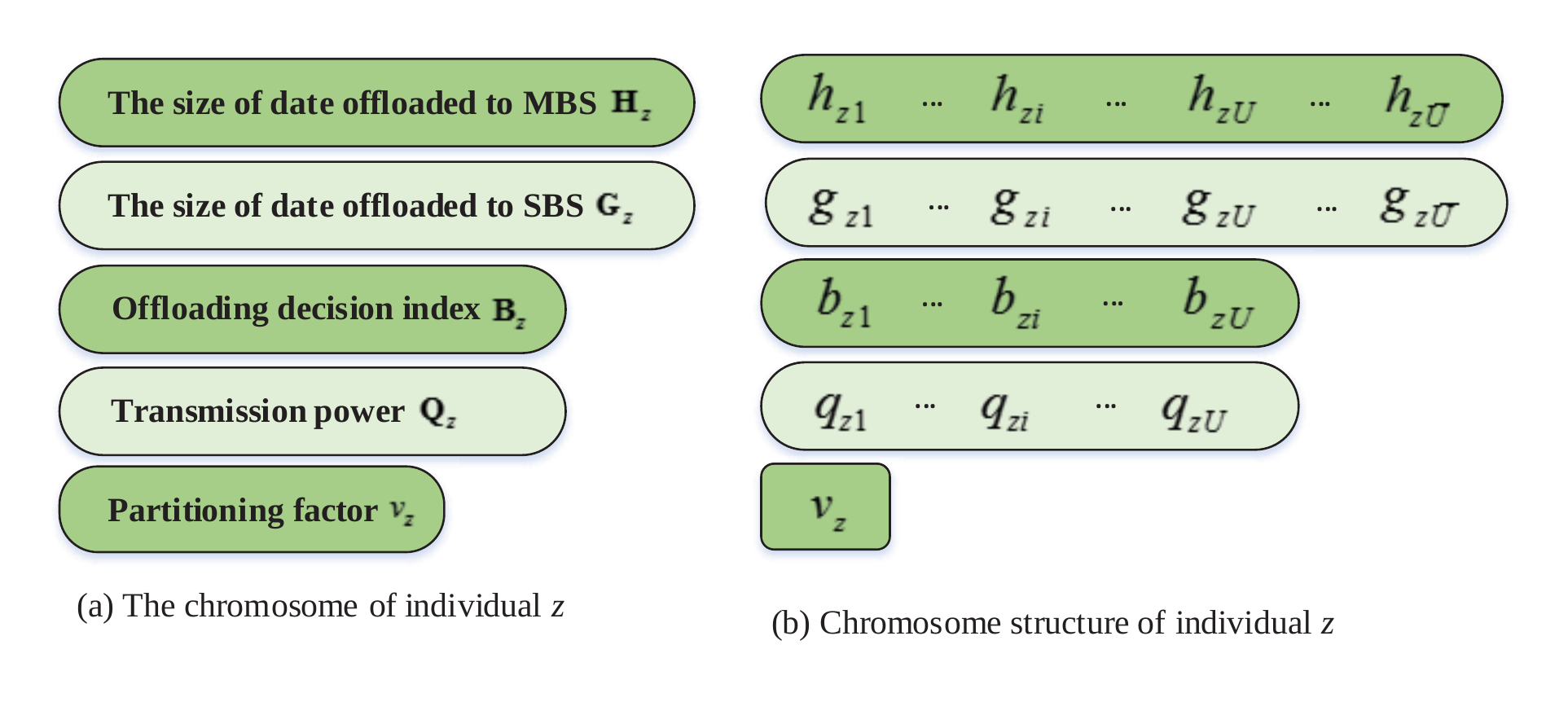}}
\caption{The chromosome and its structure of individual $z$.}
\label{fig2}
\end{figure}
\par
\textit{1) Chromosome}
\par
In GA, any individual $z$ is defined by a specific chromosome, and any chromosome can represent a solution of optimization problem. For simplicity in the design of algorithm, the real coding mechanism is utilized in this paper. Specifically, the chromosome and chromosome structure of individual $z\in \mathcal{Z}$ can be found in Fig.\ref{fig2}, where $\mathcal{Z}=\left\{ 1,2, \cdots ,Z \right\}$ represents the set of $Z$ individuals. In this figure, five optimized parameters $ \mathbf{X} $, $\mathbf{p}$, $\mathbf{\bar{D}}$, $\mathbf{\hat{D}}$, $\lambda$ are encoded as ${\mathbf{{B}}_{z}}$, ${\mathbf{{Q}}_{z}}$, ${\mathbf{{G}}_{z}}$, ${\mathbf{{H}}_{z}}$, ${{v}_{z}}$, where ${\mathbf{{B}}_{z}}=\left\{ {{b}_{zi}},i\in \mathcal{U} \right\}$, and ${{b}_{zi}}$ represents the index of BS associated with IMD $i$ in the individual $z$; $ {\mathbf{{Q}}_{z}}=\left\{ {{q}_{zi}},i\in \mathcal{U} \right\}$, and ${{q}_{zi}}$ is the transmission power of IMD $i$ in the individual $z$; $ {\mathbf{{G}}_{z}}=\left\{ {{g}_{zi}},i\in \mathcal{\bar{U}}, \right\} $, and ${{g}_{zi}}$ denotes the amount of data offloaded to associated SBS by virtual IMD $i$ in the individual $z$; $ {\mathbf{{H}}_{z}}=\left\{ {{h}_{zi}},i\in \mathcal{\bar{U}}, \right\} $, and ${{h}_{zi}}$ represents the amount of data offloaded to associated MBS by virtual IMD $i$ in the individual $z$; ${{v}_{z}}$ is the frequency band partitioning factor in individual $z$. Significantly, the set of virtual IMDs is given by $\mathcal{\bar{U}}=\left\{ 1,2, \cdots ,K, K+1,K+2, \cdots ,2K, \cdots , UK \right\} $, and $\bar{U}=UK$ is its length. Evidently, the index of any virtual IMD can be easily converted into the indices of real IMD and task. Conversely, the indices of any real IMD and task can be converted into the index of virtual IMD.
\par
\textit{2) Fitness Function}
\par
To evaluate how appropriate an individual is, a fitness function is widely used in GA. Through a direct observation, we can easily find that the constraint ${C}_{1}$ is in a nonlinear, mixed-integer and coupling form. Evidently, it may be very difficult for us to meet such constraint in the initialization and genetic operations of GA. To properly deal with the constraint ${C}_{1}$, a penalty function is introduced in the definition of fitness function, which can be definitely used for preventing individuals falling into the infeasible region. In this way, the established population can always find the feasible optimum.
\par
To minimize the network-wide energy consumption, the negative objective function of problem \eqref{eq16} can be well used as a fitness function. However, in order to meet IMDs' latency constraints at the same time, the constraint ${C}_{1}$ needs to be put into such a fitness function as a penalty term. Consequently, the fitness function of individual $z$ can be defined as
\begin{equation}\label{eq17}
	\begin{split}
		 &F\left( {\mathbf{{B}}_{z}},{\mathbf{{Q}}_{z}},{\mathbf{{G}}_{z}},{\mathbf{{H}}_{z}},{{v}_{z}} \right) \\
		 &=-E\left( {\mathbf{{B}}_{z}},{\mathbf{{Q}}_{z}},{\mathbf{{G}}_{z}},{\mathbf{{H}}_{z}},{{v}_{z}} \right)-T\left( {\mathbf{{B}}_{z}},{\mathbf{{Q}}_{z}},{\mathbf{{G}}_{z}},{\mathbf{{H}}_{z}},{{v}_{z}} \right) \\
		 &=-\sum\nolimits_{i\in \mathcal{U}}{E_{i}^{Sq}}-\sum\nolimits_{i\in \mathcal{U}}{{{\alpha }_{i}}\max\left( 0,T_{i}^{Sq}-T_{i}^{\max } \right)}, \\
	\end{split}	
\end{equation}
where $ {{\alpha }_{i}} $ is the penalty factor of IMD $i$.
\par
\textit{3) Population Initialization}
\par
To meet the constraints $C_2$-$C_7$, the initial population can be generated by using the following rules. Specifically, the initial genes of any individual $z$ can be given by
\begin{equation}\label{eq18}
	\begin{split}
		& b_{zi}^{0}=\text{randi}\left( \mathcal{S} \right),\forall i\in \mathcal{U}, \\
		& q_{zi}^{0}=\text{rand}\left( p_{i}^{\max } \right),\forall i\in \mathcal{U}, \\
		& g_{zi}^{0}=\text{rand}\left( {{d}_{mk}} \right),\forall i\in \mathcal{\bar{U}}, \\
		& h_{zi}^{0}=\text{rand}\left( g_{zi}^{0} \right),\forall i\in \mathcal{\bar{U}}, \\
		& v_{z}^{0}=\text{rand}\left( 1 \right),\\
        & \left[m,k\right]=\text{ind2sub}\left(\left[U K\right],i\right),\\
	\end{split}
\end{equation}
where $\left[m,k\right]=\text{ind2sub}\left(\left[U K\right],i\right)$ returns the row subscript $m$ and column subscript $k$ in an $U\times K$ matrix corresponding to  linear index $i$; $\text{randi}\left( \mathcal{S} \right)$ randomly outputs an element from the set $\mathcal{S}$, $ \text{rand}\left( a \right) $ randomly generates a number between 0 and $a$.
\par
\textit{4) Selection}
\par
The selection operation in GA is used for screening individuals of a population, its main task is to select some individuals from the parent population in some way and then let them be inherited into the next generation. After the selection operation, some high-fitness individuals are more likely to be inherited into the next generation, but some low-fitness ones may have a lower probability of being inherited. In this paper, the tournament method is used for selecting individuals, which should be a good option to select some good individuals rather than to select only the best individuals. Moreover, in order to improve the performance of GA, the historical best individual is always preserved in the population during the selection operation. Specifically, when the historical best individual is not chosen into the next generation, the selected worst individual should be replaced with historical best individual. In addition, the historical best individual is always updated at each iteration.
\par
\textit{5) Crossover}
\par
The crossover operation refers to the exchange of some segments of genes between two paired chromosomes according to the crossover probability, and finally establishes two new individuals. Such an operation is an important feature of GA, which is different from other evolutionary algorithms. It plays a key role in GA and is the main method used for generating new individuals. Meanwhile, it maintains the diversity of population, and it is more conducive to the convergence and avoids falling into local optimum. In this paper, we always select two neighboring individuals $z$ and $ \bar{z}=z+1$ to exchange the corresponding segments of genes. In order to propagate the building blocks of the best/better individuals \cite{MYLi2004Sep}, the crossover probability ${{{pc}}_{z\bar{z}}}$ between individuals $z$ and $ \bar{z}$ can be given by
\begin{equation}\label{eq19}
	{{{pc}}_{z\bar{z}}}=\left\{ \begin{split}
		& {{n}_{1}}\frac{{{{\bar{F}}}_{z\bar{z}}}-{{F}^{\min }}}{{{F}^{\text{ave}}}-{{F}^{\min }}},{{{\bar{F}}}_{z\bar{z}}}<{{F}^{\text{ave}}}, \\
		& {{n}_{2}},{{{\bar{F}}}_{z\bar{z}}}\ge {{F}^{\text{ave}}}, \\
	\end{split} \right.	
\end{equation}
where $ 0<{{n}_{1}}\le 1 $ and $ 0<{{n}_{2}}\le 1 $ represent the constant coefficients; ${{\bar{F}}_{z\bar{z}}}$  denotes the minimum of fitness values of individuals $z$ and $\bar{z}$. $ {{F}^{\text{ave}}} $ and ${{F}^{\min }}$ represent the average and minimum fitness values of population respectively.
\par
\textit{6) Mutation}
\par
The so-called mutation operation refers to the replacement of some gene values in individual's coding string with other gene values according to the mutation probability, and finally forms a new individual. Such an operation in GA is an auxiliary method to generate new individuals, which determines the local search ability of GA while maintaining the diversity of population. In GA, the crossover and mutation operations work together to complete global and local search of solution space. In this paper, an individual $z$ is randomly selected to mutate, and its mutation probability ${{{pm}}_{z}}$ can be given by
\begin{equation}\label{eq20}
	{{{pm}}_{z}}=\left\{ \begin{split}
		& {{n}_{3}}\frac{{{F}^{\max }}-{{{\hat{F}}}_{z}}}{{{F}^{\max }}-{{F}^{\text{ave}}}},{{{\hat{F}}}_{z}}\ge {{F}^{\text{ave}}}, \\
		& {{n}_{4}},{{{\hat{F}}}_{z}}<{{F}^{\text{ave}}}, \\
	\end{split} \right.	
\end{equation}
where $0<{{n}_{3}}<1$ and $0<{{n}_{4}}<1$ denote the constant coefficients; ${{\hat{F}}_{z}}$ represents the fitness value of individual $z$; ${{F}^{\max }}$ represents the maximum fitness value of population.
\par
Based on the mutation probability \eqref{eq20}, any chromosome of individual $z$ can be mutated by using
\begin{equation}\label{eq21}
	{{b}_{zi}}=\left\{ \begin{split}
		& \text{round}({{c}_{1}}S+\left( 1-{{c}_{1}} \right){{b}_{zi}}), \ {{c}_{2}}>0.5, \forall i\in \mathcal{U},\\
		& \text{round}\left( {{c}_{1}}+\left( 1-{{c}_{1}} \right){{b}_{zi}} \right),\ {{c}_{2}}\le 0.5, \forall i\in \mathcal{U},\\
	\end{split} \right. \\ 		
\end{equation}
\begin{equation}\label{eq22}
	{{q}_{zi}}=\left\{ \begin{split}
		& {{c}_{1}}p_{i}^{\max }+\left( 1-{{c}_{1}} \right){{q}_{zi}}, \ {{c}_{2}}>0.5,\forall i\in \mathcal{U},\\
		& \left( 1-{{c}_{1}} \right){{q}_{zi}}, \ {{c}_{2}}\le 0.5,\forall i\in \mathcal{U},\\
	\end{split} \right. \\	
\end{equation}
\begin{equation}\label{eq23}
	{{g}_{zi}}=\left\{ \begin{split}
		& {{c}_{1}}{{d}_{mk}}+\left( 1-{{c}_{1}} \right){{g}_{zi}}, \ {{c}_{2}}>0.5, \forall i\in \mathcal{\bar{U}},\\
		& \left( 1-{{c}_{1}} \right){{g}_{zi}}, \ {{c}_{2}}\le 0.5,\forall i\in \mathcal{\bar{U}},\\
	\end{split} \right. \\ 	
\end{equation}
\begin{equation}\label{eq24}
	{{h}_{zi}}=\left\{ \begin{split}
		& {{c}_{1}}{{{{g}}}_{zi}}+\left( 1-{{c}_{1}} \right){{h}_{zi}}, \ {{c}_{2}}>0.5,\forall i\in \mathcal{\bar{U}},\\
		& \left( 1-{{c}_{1}} \right){{h}_{zi}}, \ {{c}_{2}}\le 0.5,\forall i\in \mathcal{\bar{U}},\\
	\end{split} \right. \\ 	
\end{equation}
\begin{equation}\label{eq25}
	{{v}_{z}}=\left\{ \begin{split}
		& {{c}_{1}}+\left( 1-{{c}_{1}} \right){{v}_{z}}, \ {{c}_{2}}>0.5, \\
		& \left( 1-{{c}_{1}} \right){{v}_{z}}, \ {{c}_{2}}\le 0.5, \\
	\end{split} \right. \\	
\end{equation}
where $[m,k]=\text{ind2sub}([U K],i),\forall i\in \mathcal{\bar{U}}$; ${{c}_{1}}$ and ${{c}_{2}}$ are randomly generated by obeying 0-1 uniform distribution, which always keeps the optimized parameters in the feasible domain. In the rules \eqref{eq21}-\eqref{eq25}, $ {{c}_{1}} $ and ${{c}_{2}}$ are used for controlling the mutation magnitude of genes and the searching direction respectively. A larger $ {{c}_{1}} $ implies a higher mutation magnitude, $ {{c}_{2}}> 0.5 $ means some gene mutates towards the maximum, but $ {{c}_{2}}\le 0.5 $ means some gene mutates towards the minimum. It is easy to find that the mutation operation of all genes will not jump out the feasible region.
\par
In order to deal with a problem of premature convergence in traditional GA, DGM can be executed before adaptive mutation and crossover operations. To this end, we introduce the diversity measurement used for investigating the alternate between exploiting and exploring behaviors \cite{Rasmus2002Sep}. Definitely, the diversity measure for an N-dimensional numerical problem can be defined as
\begin{equation}\label{eq26}
	\begin{split}
		y=&\frac{1}{5}\Big({{\left( Z{{J}_{1}} \right)}^{-1}}\sum\nolimits_{z\in \mathcal{Z}}{\sqrt{\sum\nolimits_{i\in \mathcal{U}}{{{\left( {{b}_{zi}}-b_{i}^{\text{ave}} \right)}^{2}}}}}\Big. \\
		& +{{\left( Z{{J}_{2}} \right)}^{-1}}\sum\nolimits_{z\in \mathcal{Z}}{\sqrt{\sum\nolimits_{i\in \mathcal{U}}{{{\left( {{q}_{zi}}-q_{i}^{\text{ave}} \right)}^{2}}}}} \\
		& +{{\left( Z{{J}_{3}} \right)}^{-1}}\sum\nolimits_{z\in \mathcal{Z}}{\sqrt{\sum\nolimits_{i\in \mathcal{\bar{U}}}{{{{\left( {{g}_{zi}}-g_{i}^{\text{ave}} \right)}^{2}}}}}} \\
		& +{{\left( Z{{J}_{4}} \right)}^{-1}}\sum\nolimits_{z\in \mathcal{Z}}{\sqrt{\sum\nolimits_{i\in \mathcal{\bar{U}}}{{{{\left( {{h}_{zi}}-h_{i}^{\text{ave}} \right)}^{2}}}}}} \\
		& \Big.+{{\left( Z{{J}_{5}} \right)}^{-1}}\sum\nolimits_{z\in \mathcal{Z}}{\sqrt{{{\left( {{v}_{z}}-{{v}^{\text{ave}}} \right)}^{2}}}}\Big), \\
	\end{split}
\end{equation}
where $ {{J}_{1}} $, $ {{J}_{2}} $, $ {{J}_{3}} $, $ {{J}_{4}} $ and $ {{J}_{5}} $ represents the lengths of diagonals of feasible domains of ${\mathbf{B}}$, ${\mathbf{Q}}$, ${\mathbf{G} }$, ${\mathbf{H}}$ and ${\mathbf{v}}$ respectively;
\begin{equation}\label{eq27}
	\left\{ \begin{split}
		& b_{i}^{\text{ave}}={\sum\nolimits_{z\in \mathcal{Z}}{{{b}_{zi}}}}/{Z}, \forall i\in \mathcal{U},\\
		& q_{i}^{\text{ave}}={\sum\nolimits_{z\in \mathcal{Z}}{{{q}_{zi}}}}/{Z}, \forall i\in \mathcal{U},\\
		& g_{i}^{\text{ave}}={\sum\nolimits_{z\in \mathcal{Z}}{{{g}_{zi}}}}/{Z}, \forall i\in \mathcal{\bar{U}},\\
		& h_{i}^{\text{ave}}={\sum\nolimits_{z\in \mathcal{Z}}{{{h}_{zi}}}}/{Z}, \forall i\in \mathcal{\bar{U}},\\
		& {{v}^{\text{ave}}}={\sum\nolimits_{z\in \mathcal{Z}}{{{v}_{z}}}}/{Z}. \\
	\end{split} \right.
\end{equation}
Then, DGM can be executed according to a specified probability, i.e.,
\begin{equation}\label{eq28}
	\text{pd=}\left\{ \begin{split}
		& {{n}_{5}},\ if\ y<{{y}_{1}}, \\
		& {{n}_{6}},\ if\ {{y}_{1}}\le y<{{y}_{2}}, \\
		& {{n}_{7}},\ \text{otherwise}, \\
	\end{split} \right.
\end{equation}
where $0<y_1<1$ and $0<y_2<1$ are constant coefficients, and represent the diversity thresholds.
\par
Up to now, adaptive GA with DGM can be summarized in Algorithm 1, where $T_1$ denotes the number of iterations.
\begin{table}[]
	\centering
	\begin{tabular}{ll}
		\toprule[1pt]
		\textbf{Algorithm 1: Adaptive GA with DGM} \\ \midrule[0.5pt]
		1: \textbf{Initialization:}\\
        2:\ \ \ \ $t_1=1$.\\
		3:\ \ \ \ Initialize the population including $Z$ individuals using \eqref{eq18}.\\
		4:\ \ \ \ Calculate the fitness values of all individuals using \eqref{eq17}.\\
		5:\ \ \ \ Find the current best individual with maximum fitness values.\\
		6:\ \ \ \ Replace historical best individual with current best individual if\\
		7:\ \ \ \ \ \ the former individual has smaller fitness value than the later one.\\
		8: \textbf{While $t_1<=T_1$} \textbf{do}\\
		9:\ \ \ \ Generate a new population including $Z$ individuals selected by \\
		10:\ \ \ \ \ \ tournament method.\\
		11:\ \ \ Replace worst individual with historical best individual If the \\
		12:\ \ \ \ \ latter is not selected into the next generation. \\
		13:\ \ \ Execute DGM using \eqref{eq21}-\eqref{eq25} under the probability
            \eqref{eq28}.\\
		14:\ \ \ Calculate the fitness values of all individuals using \eqref{eq17}.\\
		15:\ \ \ Adaptively cross any two neighbouring individuals under \eqref{eq19}.\\
        16:\ \ \ Adaptively mutate using \eqref{eq21}-\eqref{eq25} under the probability \eqref{eq20}.\\
		17:\ \ \ Calculate the fitness values of all individuals using \eqref{eq17}.\\
        18:\ \ \ Find the current best individual in this generation.\\
		19:\ \ \ Replace historical best individual with current best individual if\\
		20:\ \ \ \ \ the former individual has smaller fitness value than the later one.\\
		21:\ \ \ Update the iteration index: ${t_1}={t_1}+1$.\\
		22: \textbf{EndWhile}\\ \bottomrule[0.5pt]
	\end{tabular}
	\label{alg1}
\end{table}
\subsection{Adaptive PSO}
As a population-based intelligent optimization algorithm, PSO algorithm was proposed according to the foraging behavior of birds. In the PSO, each particle has two attributes including position and velocity, where the position represents a solution of optimization problem, and the velocity shows how the solutions evolve. Specifically, the velocity of any particle (individual) $z$ can be updated by
\begin{equation}\label{eq29}
	\begin{split}
		bv_{zi}^{t+1}=
		& \kappa _{z}^{t}bv_{zi}^{t}+{{w}_{1}}{{\eta }_{zi}}\left( bpbest_{zi}^{t}-bl_{zi}^{t} \right) \\	
		&+{{w}_{2}}{{{\hat{\eta }}}_{zi}}(bgbest_{i}^{t}-bl_{zi}^{t}), \forall i\in \mathcal{U}, \\
	\end{split}
\end{equation}
\begin{equation}\label{eq30}
	\begin{split}	
		qv_{zi}^{t+1}=
		& \kappa _{z}^{t}qv_{zi}^{t}+{{w}_{1}}{{\eta }_{zi}}\left( qpbest_{zi}^{t}-ql_{zi}^{t} \right)\\
		& +{{w}_{2}}{{{\hat{\eta }}}_{zi}}(qgbest_{i}^{t}-ql_{zi}^{t}), \forall i\in \mathcal{U}, \\
	\end{split}
\end{equation}
\begin{equation}\label{eq31}
	\begin{split}
		gv_{zi}^{t+1}=
		& \kappa _{z}^{t}gv_{zi}^{t}+{{w}_{1}}{{\mu }_{zi}}\left( gpbest_{zi}^{t}-gl_{zi}^{t} \right)\\
		& +{{w}_{2}}{{{\hat{\mu }}}_{zi}}(ggbest_{i}^{t}-gl_{zi}^{t}), \forall i\in \mathcal{\bar{U}}, \\
	\end{split}
\end{equation}
\begin{equation}\label{eq32}
	\begin{split}
		hv_{zi}^{t+1}=
		&\kappa _{z}^{t}hv_{zi}^{t}+{{w}_{1}}{{\mu }_{zi}}\left( hpbest_{zi}^{t}-hl_{zi}^{t} \right)\\
		&+{{w}_{2}}{{{\hat{\mu }}}_{zi}}(hgbest_{i}^{t}-hl_{zi}^{t}), \forall i\in \mathcal{\bar{U}}, \\
	\end{split}
\end{equation}
\begin{equation}\label{eq33}
	\begin{split}
		vv_{z}^{t+1}=
		& \kappa _{z}^{t}vv_{z}^{t}+{{w}_{1}}{{\gamma }_{z}}\left( vpbest_{z}^{t}-vl_{z}^{t} \right)\\
		& +{{w}_{2}}{{{\hat{\gamma }}}_{z}}({{vgbest}^{t}}-vl_{z}^{t}), \\
	\end{split}
\end{equation}
where $bv_{zi}^{t}$, $qv_{zi}^{t}$, $gv_{zi}^{t} $, $ hv_{zi}^{t} $ and $vv_{z}^{t}$ are the velocities of ${{b}_{zi}}$, ${{q}_{zi}}$, ${{g}_{zi}}$, ${{h}_{zi}}$ and ${{v}_{z}}$ at $t$-th iteration; $bl_{zi}^{t}$, $ql_{zi}^{t}$, $gl_{zi}^{t} $, $ hl_{zi}^{t} $ and $vl_{z}^{t}$ are the positions of ${{b}_{zi}}$, ${{q}_{zi}}$, ${{g}_{zi}}$, ${{h}_{zi}}$ and ${{v}_{z}}$ at $t$-th iteration; $\kappa _{z}^{t}$ represents an inertia weight of particle $z$ at the $t$-th iteration; $0<{{w}_{1}}<1$ and $0<{{w}_{2}}<1$ denote the self-learning and social learning factors respectively; $\eta_{zi}$, $\hat{\eta}_{zi}$, $\mu_{zi}$, $\hat{\mu}_{zi}$, $\gamma_{z}$ and $\hat{\gamma}_{z}$ are random numbers; $\mathbf{bpbest}_{z}^{t}=\{ {{ bpbest}_{zi}^{t}},\forall i\in \mathcal{U} \}$, $\mathbf{qpbest}_{z}^{t}=\{ {{ qpbest}_{zi}^{t}},\forall i\in \mathcal{U} \}$, $\mathbf{gpbest}_{z}^{t}=\{ {{ gpbest}_{zi}^{t}},\forall i\in \bar{\mathcal{U}} \}$, $\mathbf{hpbest}_{z}^{t}=\{ {{ hpbest}_{zi}^{t}},\forall i\in \bar{\mathcal{U}} \}$ and ${{ vpbest}_{z}^{t}}$ are the historical optimal position of particle $z$ at the $t$-th iteration; $\mathbf{bgbest}^{t}=\{ {{ bgbest}_{i}^{t}},\forall i\in \mathcal{U} \}$, $\mathbf{qgbest}^{t}=\{ {{ qgbest}_{i}^{t}},\forall i\in \mathcal{U} \}$, $\mathbf{ggbest}^{t}=\{ {{ ggbest}_{i}^{t}},\forall i\in \bar{\mathcal{U}} \}$, $\mathbf{hgbest}^{t}=\{ {{ hgbest}_{i}^{t}},\forall i\in \bar{\mathcal{U}} \}$ and ${{ vgbest}^{t}}$ are the global historical optimal position, i.e., the position of global historical best particle at the $t$-th iteration. In this paper, the particle who has historical optimal position is regarded as personal best particle, and the one who has global historical optimal position is seen as global best particle. Evidently, the global best particle is selected from $Z$ personal best particles.
\par
In \eqref{eq29}-\eqref{eq33}, the inertia weight $\kappa _{z}$ of any individual $z$ can be updated by
\begin{equation}\label{eq34}
	\kappa _{z}^{t+1}=\left\{ \begin{split}
		& \kappa _{z}^{t}+{t\left( {{\kappa }^{\max }}-{{\kappa }^{\min }} \right)}/{T_2}\;,\text{for best particle}, \\
		& \kappa _{z}^{t}-{t\left( {{\kappa }^{\max }}-{{\kappa }^{\min }} \right)}/{T_2}\;,\text{otherwise}, \\
	\end{split} \right.
\end{equation}
where $ {{\kappa }^{\max }} $ and $ {{\kappa }^{\min }} $  represent the maximum and minimum inertia weights respectively; $T_2$ is the number of iterations of adaptive PSO.
\par
After updating the velocities of particles, the position of any particle $z$ can be updated by
\begin{equation}\label{eq35}
	bl_{zi}^{t+1}=\text{round}\left( bl_{zi}^{t}+bv_{zi}^{t+1} \right), \forall i\in \mathcal{U},  \\
\end{equation}
\begin{equation}\label{eq36}
	ql_{zi}^{t+1}=ql_{zi}^{t}+qv_{zi}^{t+1}, \forall i\in \mathcal{U},  \\
\end{equation}
\begin{equation}\label{eq37}
	gl_{zi}^{t+1}=gl_{zi}^{t}+gv_{zi}^{t+1}, \forall i\in \mathcal{\bar{U}},  \\
\end{equation}
\begin{equation}\label{eq38}
	hl_{zi}^{t+1}=hl_{zi}^{t}+hv_{zi}^{t+1}, \forall i\in \mathcal{\bar{U}},  \\
\end{equation}
\begin{equation}\label{eq39}
	vl_{z}^{t+1}=vl_{z}^{t}+vv_{z}^{t+1}. \\
\end{equation}
\par
In order to keep global best particle $\hat{z}$ moving toward the local minimum, according to the rules in \cite{FVanddenbergh2007}, the velocity of particle $\hat{z}$ is updated again by
\begin{equation}\label{eq40}
\begin{split}
bv_{\hat{z}i}^{t+1}=&-bl_{\hat{z}i}^{t}+bgbest_{i}^{t}+{{w}_{3}}bv_{\hat{z}i}^{t}\\
              &+{{\beta }^{t}}\left( 1-2{{\delta }_{\hat{z}i}} \right), \forall i\in \mathcal{U}, \\
\end{split}
\end{equation}
\begin{equation}\label{eq41}
\begin{split}
qv_{\hat{z}i}^{t+1}=&-ql_{\hat{z}i}^{t}+qgbest_{i}^{t}+{{w}_{3}}qv_{\hat{z}i}^{t}\\
&+{{\beta }^{t}}\left( 1-2{{\delta }_{\hat{z}i}} \right), \forall i\in \mathcal{U}, \\
\end{split}
\end{equation}
\begin{equation}\label{eq42}
\begin{split}
gv_{\hat{z}i}^{t+1}=&-gl_{\hat{z}i}^{t}+ggbest_{i}^{t}+{{w}_{3}}gv_{\hat{z}i}^{t}\\
&+{{\beta }^{t}}\left( 1-2{\bar{\delta }_{\hat{z}i}} \right), \forall i\in \mathcal{\bar{U}},  \\
\end{split}	
\end{equation}
\begin{equation}\label{eq43}
\begin{split}
hv_{\hat{z}i}^{t+1}=&-hl_{\hat{z}i}^{t}+hgbest_{i}^{t}+{{w}_{3}}hv_{\hat{z}i}^{t}\\
&+{{\beta }^{t}}\left( 1-2{\bar{\delta }_{\hat{z}i}} \right), \forall i\in \mathcal{\bar{U}},  \\
\end{split}		
\end{equation}
\begin{equation}\label{eq44}
\begin{split}
vv_{\hat{z}}^{t+1}=&-vl_{\hat{z}}^{t}+{{vgbest}^{t}}+{{w}_{3}}vv_{\hat{z}}^{t}\\
&+{{\beta }^{t}}\left( 1-2{\hat{\delta }_{\hat{z}}} \right), \\
\end{split}		
\end{equation}
where ${{\beta }^{t}}$ denotes a scaling factor at $t$-th iteration; ${{w}_{3}}$ is a constant coefficient; $0\le {{\delta }_{\hat{z}i}}\le 1$, $0\le {\bar{\delta }_{\hat{z}i}}\le 1$ and $0\le {\hat{\delta }_{\hat{z}}}\le 1$ are random numbers.
\par
Then, according to the rules in \cite{FVanddenbergh2007}, the position of global best particle $\hat{z}$ is updated again by
\begin{equation}\label{eq45}
	bl_{\hat{z}i}^{t+1}=\text{round}\left( bgbest_{i}^{t}+{{w}_{3}}bv_{\hat{z}i}^{t}+{{\beta }^{t}}\left( 1-2{{\delta }_{\hat{z}i}} \right) \right), \forall i\in \mathcal{U}, \\
\end{equation}
\begin{equation}\label{eq46}
	ql_{\hat{z}i}^{t+1}=qgbest_{i}^{t}+{{w}_{3}}qv_{\hat{z}i}^{t}+{{\beta }^{t}}\left( 1-2{{\delta }_{\hat{z}i}} \right), \forall i\in \mathcal{U}, \\
\end{equation}
\begin{equation}\label{eq47}
	gl_{\hat{z}i}^{t+1}=ggbest_{i}^{t}+{{w}_{3}}gv_{\hat{z}i}^{t}+{{\beta }^{t}}\left( 1-2{\bar{\delta }_{\hat{z}i}} \right), \forall i\in \mathcal{\bar{U}}, \\
\end{equation}
\begin{equation}\label{eq48}
	hl_{\hat{z}i}^{t+1}=hgbest_{\hat{z}i}^{t}+{{w}_{3}}hv_{\hat{z}i}^{t}+{{\beta }^{t}}\left( 1-2{\bar{\delta }_{\hat{z}i}} \right), \forall i\in \mathcal{\bar{U}}, \\
\end{equation}
\begin{equation}\label{eq49}
	vl_{\hat{z}}^{t+1}={{vgbest}^{t}}+{{w}_{3}}vv_{\hat{z}}^{t}+{{\beta }^{t}}\left( 1-2{\hat{\delta }_{\hat{z}}} \right), \\
\end{equation}
where scaling factor $\beta$, used for driving PSO to randomly search the surrounding area of global best position $\mathbf{gbest}=\{\mathbf{bgbest}, \mathbf{qgbest}, \mathbf{ggbest},\mathbf{hgbest},vgbest\}$, can be updated by
\begin{equation}\label{eq50}
	{{\beta }^{t+1}}=\left\{ \begin{split}
		& 2{{\beta }^{t}},\ {{{N}}_{1}}>{{\mathfrak{u}}_{1}}, \\
		& 0.5{{\beta }^{t}},\ {{{N}}_{2}}>{{\mathfrak{u}}_{2}}, \\
		& {{\beta }^{t}},\ \text{otherwise}, \\
	\end{split} \right.	
\end{equation}
$ {{N}_{1}} $ is the number of consecutive successes, $ {{N}_{2}} $ is the one of consecutive failures, $ {{\mathfrak{u}}_{1}} $ and $ {{\mathfrak{u}}_{2}} $ represent the threshold parameters. It is worth while to note that $F(\mathbf{gbest}^{t})=F(\mathbf{gbest}^{t-1})$ shows the failure of search, but the other cases imply the successes.
\begin{table}[]
	\centering
	\begin{tabular}{ll}
		\toprule[1pt]
		\textbf{Algorithm 2: Adaptive PSO} \\ \midrule[0.5pt]
		1: \textbf{Initialization:}\\
        2:\ \ \ \ $t_2=1$.\\
        3:\ \ \ \ Initialize the velocities of all particles.\\
		4:\ \ \ \ Initialize the velocities of personal best particles.\\
        5:\ \ \ \ Initialize the positions of all particles.\\
		6:\ \ \ \ Find the global best particle of current population.\\
		7: \textbf{While $t_2<=T_2$} \textbf{do}\\
		8:\ \ \ \ Update inertia weight using \eqref{eq34}.\\
		9:\ \ \ \ Update the velocities of all particles using \eqref{eq29}-\eqref{eq33}.\\
		10:\ \ \ Update the positions of all particles using \eqref{eq35}-\eqref{eq39}.\\
		11:\ \ \ Calculate the fitness values of all particles using \eqref{eq17}.\\
		12:\ \ \ For any particle, the velocity and position of personal best particle \\
		13:\ \ \ \ is updated using its current attributes if the current fitness value\\
		14:\ \ \ \ of this particle is higher than its historical optimal one.\\
		15:\ \ \ Find the global best particle among personal best particles.\\
		16:\ \ \ Update the velocity of global best particle using \eqref{eq40}-\eqref{eq44}.\\
		17:\ \ \ Update the position of global best particle using \eqref{eq45}-\eqref{eq49}.\\
		18:\ \ \ Update the factor $\beta$ using \eqref{eq50}.\\
		19:\ \ \ Update the iteration index: $t_2=t_2+1$.\\
		20: \textbf{EndWhile}\\ \bottomrule[0.5pt]
	\end{tabular}
	\label{alg2}
\end{table}
\par
Up to now, the whole procedure developed for adaptive PSO can be summarized in Algorithm 2, where $T_2$ is the number of iterations. Then, the whole procedure designed for solving the problem \eqref{eq16} can be summarized in Algorithm 3. In such an algorithm, Algorithm 1 is used for coarse-grained search at first, and then Algorithm 2 is utilized for fine-grained search. Evidently, it just is a one-layer iterative algorithm, and can be well implemented in the reality.
\begin{table}[]
	\centering
	\begin{tabular}{ll}
		\toprule[1pt]
		\textbf{Algorithm 3: Hierarchical Adaptive Search (HAS)} \\ \midrule[0.5pt]
	1: Enter initialization parameters.\\
	2: Search a coarse-grained solution using Algorithm 1.\\
	3: Search a fine-grained solution using Algorithm 2.\\
	 \bottomrule[0.5pt]
	\end{tabular}
	\label{alg3}
\end{table}
\section{Algorithm Analysis}\label{sec 5}
In this section, the convergence, computation complexity and parallel implement of aforementioned algorithms will be made.
\subsection{Convergence analysis}
By following the procedure in \cite{TZou2021Oct}, the convergence of adaptive GA with DGM and adaptive PSO can be established as follows.
\par
\noindent
\textbf{\textit{Theorem 1:}} Adaptive GA with DGM is globally convergent.
\par
\textit{Proof:}  As revealed in \cite{MYLi2004Sep}, if GA can always maintains the best individual after/before selection, it will converge to the global optimum. In adaptive GA with DGM, a tournament method is utilized to select $Z$ individuals, and the historical best individual takes place of current worst one after the selection operation. In other words, the historical best individual can be always maintained after selection. Consequently, adaptive GA with DGM can converge to the global optimum. \ding{113}
\par
When the DGM probabilities ${{n}_{5}}=0$, ${{n}_{6}}=0$ and ${{n}_{7}}=0$ exist in \eqref{eq28}, adaptive GA with DGM just is a traditional adaptive GA. Then, the following results can be easily established for Theorem 1.
\par
\noindent
\textit{\textbf{Corollary 1:}} Adaptive GA without DGM converges to the global optimum if it always maintains best individual after/before selection operation.
\par
Next, we will investigate the convergence of adaptive PSO. By following the convergence proving in \cite{FVanddenbergh2007}, some important results can be summarized as follows.
\par
\noindent
\textit{\textbf{Lemma 1:}} Adaptive PSO meets the condition H1 in \cite{FVanddenbergh2007}.
\par
\textit{Proof}: At first, a function ${{\mathcal{F}}_{1}}$ is defined as
\begin{equation}\label{eq51}
	{{\mathcal{F}}_{1}}\left( {{\mathbf{{gbest}}}^{t}},\mathbf{lc}_{z}^{t} \right)=\left\{ \begin{split}
		& {{\mathbf{{gbest}}}^{t}},\ \text{if}\ F\left(\mathbf{lc}_{z}^{t+1} \right)\le F\left(  {{\mathbf{{gbest}}}^{t}} \right), \\
		& \mathbf{lc}_{z}^{t+1},\ \text{otherwise}, \\
	\end{split} \right.	
\end{equation}	
where $\mathbf{lc}_{z}^{t+1}=\left\{ \mathbf{bl}_{z}^{t+1},\mathbf{ql}_{z}^{t+1},\mathbf{ gl}_{z}^{t+1},\mathbf{hl}_{z}^{t+1},{vl}_{z}^{t+1} \right\}$;
\begin{equation}\label{eq52}
	\mathbf{bl}_{z}^{t+1}=\text{round}\left(\begin{split} &\mathbf{bl}_{z}^{t}+\kappa _{z}^{t}\mathbf{bv}_{z}^{t}+{{w}_{1}}{\boldsymbol{\eta }_{z}}\left( \mathbf{bpbest}_{z}^{t}-\mathbf{bl}_{z}^{t}\right)\\
		&+{{w}_{2}}{{\boldsymbol{\hat{\eta }}}_{z}}\left(\mathbf{bgbest}^{t}-\mathbf{bl}_{z}^{t}\right),
	\end{split} \right) \\
\end{equation}
\begin{equation}\label{eq53}
	\begin{split}
	\mathbf{ql}_{z}^{t+1}=
	&\mathbf{ql}_{z}^{t}+\kappa _{z}^{t}\mathbf{qv}_{z}^{t}+{{w}_{1}}{\boldsymbol{\eta }_{z}}\left( \mathbf{qpbest}_{z}^{t}-\mathbf{ql}_{z}^{t} \right)\\
	&+{{w}_{2}}{{\boldsymbol{\hat{\eta }}}_{z}}(\mathbf{qgbest}^{t}-\mathbf{ql}_{z}^{t}),
\end{split}
\end{equation}
\begin{equation}\label{eq54}
	\begin{split}
		\mathbf{gl}_{z}^{t+1}=
		&\mathbf{gl}_{z}^{t}+\kappa _{z}^{t}\mathbf{gv}_{z}^{t}+{{w}_{1}}{\boldsymbol{\mu }_{z}}\left( \mathbf{gpbest}_{z}^{t}-\mathbf{gl}_{z}^{t} \right)\\
		&+{{w}_{2}}{{\boldsymbol{\hat{\mu }}}_{z}}(\mathbf{ggbest}^{t}-\mathbf{gl}_{z}^{t}),
	\end{split}
\end{equation}
\begin{equation}\label{eq55}
	\begin{split}
		\mathbf{hl}_{z}^{t+1}=
		&\mathbf{hl}_{z}^{t}+\kappa _{z}^{t}\mathbf{hv}_{z}^{t}+{{w}_{1}}{\boldsymbol{\mu }_{z}}\left( \mathbf{hpbest}_{z}^{t}-\mathbf{hl}_{z}^{t} \right)\\
		&+{{w}_{2}}{{\boldsymbol{\hat{\mu }}}_{z}}(\mathbf{hgbest}^{t}-\mathbf{hl}_{z}^{t}),
	\end{split}
\end{equation}
\begin{equation}\label{eq56}
	\begin{split}
		{vl}_{z}^{t+1}=
		&{vl}_{z}^{t}+\kappa _{z}^{t}{vv}_{z}^{t}+{{w}_{1}}{{\gamma }_{z}}\left( {vpbest}_{z}^{t}-{vl}_{z}^{t} \right)\\
		&+{{w}_{2}}{{{\hat{\gamma }}}_{z}}({vgbest}^{t}-{vl}_{z}^{t}).
	\end{split}
\end{equation}
In \eqref{eq52}-\eqref{eq56}, ${\boldsymbol{\eta }_{z}}=\{ {{\eta}_{zi}},\forall i\in \mathcal{U} \}$, ${\boldsymbol{\hat{\eta} }_{z}}=\{ {\hat{\eta}_{zi}},\forall i\in \mathcal{U} \}$, ${\boldsymbol{\mu }_{z}}=\{ {{\mu}_{zi}},\forall i\in \bar{\mathcal{U}} \}$ and ${\boldsymbol{\hat{\mu} }_{z}}=\{ {\hat{\mu}_{zi}},\forall i\in \bar{\mathcal{U}} \}$.
\par
Then, we can easily achieve
\begin{equation}\label{eq57}
	F\left[ {{\mathcal{F}}_{1}}\left( \mathbf{gbest}^{t},\mathbf{lc}_{z}^{t} \right) \right]\ge F\left( \mathbf{gbest}^{t} \right),
\end{equation}
Evidently, the conventional PSO consisting of \eqref{eq29}-\eqref{eq39} meets the condition H1 in \cite{FVanddenbergh2007}.
\par
Next, we study the update \eqref{eq40}-\eqref{eq49}. To this end, we firstly define another function $\mathcal{F}_{2}$, which is given by
\begin{equation}\label{eq58}
	{{\mathcal{F}}_{2}}\left( \mathbf{gbest}^{t},\mathbf{lc}_{\hat{z}}^{t} \right)=\left\{ \begin{split}
		& \mathbf{gbest}^{t},\ \text{if}\ F\left( \mathbf{lc}_{\hat{z}}^{t+1} \right)\le F\left(\mathbf{gbest}^{t} \right), \\
		& \mathbf{lc}_{\hat{z}}^{t+1}, \text{otherwise}, \\
	\end{split} \right.
\end{equation}
where $\mathbf{lc}_{\hat{z}}^{t+1}=\left\{ \mathbf{bl}_{\hat{z}}^{t+1},\mathbf{ql}_{\hat{z}}^{t+1},\mathbf{ gl}_{\hat{z}}^{t+1},\mathbf{hl}_{\hat{z}}^{t+1},{vl}_{\hat{z}}^{t+1} \right\}$ represents the position of global best particle $\hat{z}$;
\begin{equation}\label{eq59}
	\mathbf{bl}_{\hat{z}}^{t+1}=\text{round}\left( \mathbf{bgbest}^{t}+{{w}_{3}}\mathbf{bv}_{\hat{z}}^{t}+{{\beta }^{t}}\left( 1-2{\boldsymbol{\delta }_{\hat{z}}} \right) \right),
\end{equation}
\begin{equation}\label{eq60}
	\mathbf{ql}_{\hat{z}}^{t+1}=\mathbf{qgbest}^{t}+{{w}_{3}}\mathbf{qv}_{\hat{z}}^{t}+{{\beta }^{t}}\left( 1-2{\boldsymbol{\delta }_{\hat{z}}} \right),
\end{equation}
\begin{equation}\label{eq61}
	\mathbf{gl}_{\hat{z}}^{t+1}=\mathbf{ggbest}^{t}+{{w}_{3}}\mathbf{gv}_{\hat{z}}^{t}+{{\beta }^{t}}\left( 1-2{\boldsymbol{\bar{\delta }}_{\hat{z}}} \right),
\end{equation}
\begin{equation}\label{eq62}
	\mathbf{hl}_{\hat{z}}^{t+1}=\mathbf{hgbest}^{t}+{{w}_{3}}\mathbf{hv}_{\hat{z}}^{t}+{{\beta }^{t}}\left( 1-2{\boldsymbol{\bar{\delta }}_{\hat{z}}} \right),
\end{equation}
\begin{equation}\label{eq63}
	{vl}_{\hat{z}}^{t+1}={vgbest}^{t}+{{w}_{3}}{vv}_{\hat{z}}^{t}+{{\beta }^{t}}\left( 1-2{\hat{\delta }_{\hat{z}}} \right).
\end{equation}
In \eqref{eq59}-\eqref{eq63},  ${\boldsymbol{\delta }_{\hat{z}}}=\{ {{\delta}_{\hat{z}i}},\forall i\in \mathcal{U} \}$ and ${\boldsymbol{\bar{\delta} }_{\hat{z}}}=\{ {\bar{\delta}_{\hat{z}i}},\forall i\in \mathcal{\bar{U}} \}$.
\par
Then, we can easily achieve
\begin{equation}\label{eq64}
	F\left[ {{\mathcal{F}}_{2}}\left( \mathbf{gbest}^{t},\mathbf{lc}_{\hat{z}}^{t} \right) \right]\ge F\left( \mathbf{gbest}^{t} \right).
\end{equation}
Based on the analyses mentioned above, it is evident that adaptive PSO meets the condition H1 in \cite{FVanddenbergh2007}. \ding{113}
\par
\noindent
\textbf{\textit{Lemma 2:}} Adaptive PSO meets the condition H3 in  \cite{FVanddenbergh2007}.
\par
\textit{Proof:} According to the results in \cite{FVanddenbergh2007}, some similar deductions for adaptive PSO used in this paper can be made as follows. To this end, we give the definitions of $\mathcal{L}_{0}$ and $ \mathcal{L}_{\epsilon}$ at first. For any $ \mathbf{W}_{0}\in \mathcal{R} $, $ \mathcal{L}_{0}=\left\{\mathbf{W}\in\mathcal{R} \left| F\left( \mathbf{W} \right) \right.\ge F\left( {{\mathbf{W}}_{0}} \right) \right\}$, where $ \mathcal{R} $ denotes the feasible region of $\mathbf{lc}=\left\{ \mathbf{bl},\mathbf{ql},\mathbf{ gl},\mathbf{hl},\mathbf{vl} \right\} $. In addition, $ \mathcal{L}_{\epsilon}=\left\{\mathbf{W}\in\mathcal{R} \left| F\left( \mathbf{W} \right) \right.\ge \chi+\epsilon \right\} $, where $\epsilon\textgreater{0}$, $\chi=\sup\left( \mathcal{A}:\mathcal{Y}\left[ \mathbf{W}\in \mathcal{R}\left| F\left( \mathbf{W} \right)>\mathcal{A} \right. \right]>0\right) $ denotes the essential supremum of $F\left(\mathbf{W}\right)$ under Lebesgue measure $\mathcal{Y}\left[ \mathbf{W}\in \mathcal{R}\left| F\left( \mathbf{W} \right)>\mathcal{A} \right. \right]>0$.
\par
We assume that $\mathbf{gbest}$ is in $ \mathcal{L}_{0} $. According to the rules of adaptive PSO, we can easily know that it always generates a sample around some point in $ \mathcal{L}_{0} $. Moreover, adaptive PSO starts with any point in $\mathcal{L}_{0}$, it can always guarantee a non-degenerate sampling volume in a nonzero probability of sampling a point closer to the optimality region $ \mathcal{L}_{\epsilon} $. Based on these results, we can easily know that adaptive PSO meets the condition H3 in \cite{FVanddenbergh2007}. \ding{113}
\par
\noindent
\textbf{\textit{Theorem 2:}} Adaptive PSO converges to the local minimum.
\par
\textit{Proof:} As revealed in \cite{FVanddenbergh2007}, since adaptive PSO  meets the conditions H1 and H3 in \cite{FVanddenbergh2007}, it can converges to the local minimum finally. \ding{113}
\subsection{Complexity Analysis}
The analyses of computation complexity of adaptive GA with DGM, adaptive PSO and HAS can be summarized as follows.
\par
\noindent
\textbf{\textit{Proposition 1}}: The computation complexity of adaptive GA with DGM is $\mathcal{O}\left(KUZS{T_1}\right)$ after $T_1$ iterations.
\par
\textit{Proof:} In this paper, the tournament method is utilized to generate new population. In each round of this method, we assume that two individuals are selected. As revealed in \cite{FGuo2018Dec}, the selection operation in adaptive GA with DGM achieves a complexity of $\mathcal{O}\left( {2Z} \right)$. In addition, for the chromosomes whose length is $U$, the crossover operation has a complexity of $\mathcal{O}\left( {3U} \right)$, and the mutation operation has a complexity of $\mathcal{O}\left({US} \right)$. For the chromosomes whose length is $KU$, the crossover operation has a complexity of $\mathcal{O}\left( {3KU} \right)$, and the mutation operation has a complexity of $\mathcal{O}\left({KUS} \right)$.
\par
In the Step 3 of Algorithm 1 (adaptive GA with DGM), the conversion between subscript and index can be executed before the population initialization, which has a complexity of $\mathcal{O}\left( KU \right)$ for each individual. Then, the computation complexity of Step 3 in Algorithm 1 is $\mathcal{O}\left(KUZ\right)$ for $Z$ individuals. In the Steps 4, 14 and 17 of Algorithm 1, we can calculate IMDs' data rates and computation capacities before executing these steps. The former calculation has a complexity of $\mathcal{O}\left(US\right)$  for each individual, but the latter one has a complexity of $\mathcal{O}\left(KUS\right)$  for each individual. Therefore, Steps 4, 14 and 17 achieve a complexity of $\mathcal{O}\left(KUZS\right)$ for $Z$ individuals. In addition, Steps 5 and 18 have a complexity of $\mathcal{O}\left(Z\right)$, Steps 6-7, 11-12, and 19-20 have a complexity of $\mathcal{O}\left(KU\right)$.
\par
In general, the computation complexity of adaptive GA with DGM is $\mathcal{O}\left(KUZS{T_1}\right)$ after $T_1$ iterations. \ding{113}
\par
\noindent
\textbf{\textit{Proposition 2}}: The computation complexity of adaptive PSO is $\mathcal{O}\left(KUZS{T_2} \right)$ after ${{T}_{2}}$ iterations.
\par
\emph{Proof}: In Steps 3-6 of Algorithm 2 (Adaptive PSO), the computation complexity is mainly derived from the initialization of ${gl}_{zi}$, ${hl}_{zi}$, ${gv}_{zi}$ and ${hv}_{zi}$ for all $z$ and $i$, which is $\mathcal{O}\left(KUZ \right)$. In Step 8, the computation complexity is $\mathcal{O}\left(Z \right)$. In Steps 9-11,  the computation complexity is mainly derived from the update of ${gl}_{zi}$, ${hl}_{zi}$, ${gv}_{zi}$ and ${hv}_{zi}$ for all $z$ and $i$, which is $\mathcal{O}\left(KUZ \right)$. In Step 11, the computation complexity is mainly derived from the calculation of fitness values of all individuals, which is $\mathcal{O}\left(KUZS \right)$. In addition, it is easy to find that the computation complexity of Steps 12-14 is $\mathcal{O}\left(KU \right)$, the one of Step 15 is $\mathcal{O}\left(Z \right)$, and the one of Steps 16-17  is $\mathcal{O}\left(KU\right)$.
\par
It is evident that the computation complexity of Algorithm 2 is mainly derived from the calculation of fitness values of all individuals in Step 11. Therefore, the computation complexity of adaptive PSO is $\mathcal{O}\left(KUZS{T_2} \right)$ after $T_2$ iterations. \ding{113}
\par
\noindent
\textbf{\textit{Proposition 3}}: The computation complexity of HAS is $\mathcal{O}\left(KUZS\bar{T}\right)$, where $\bar{T}=\max({T_1},{T_2})$.
\par
\textit{Proof:} In HAS, we execute adaptive GA with DGM and adaptive PSO sequentially. Therefore, the computation complexity of HAS is the maximum between $\mathcal{O}\left(KUZS{T_1}\right)$ and $\mathcal{O}\left(KUZS{T_2} \right)$.
\subsection{Parallel Implement Analysis}
Through a direct observation on adaptive GA with DGM and adaptive PSO, we can easily find that the computation complexity of them is mainly derived from the calculation of fitness values of all individuals. Such a calculation will result in a relatively high computation complexity if the number of individuals is too large. To reduce the computation complexity and improve the efficiency of algorithms, all individuals can definitely calculate their fitness values in a parallel manner, which has been widely advocated in the reality.
\section{NUMERICAL RESULTS}\label{sec 6}
Without loss of generality, IMDs and ultra-dense SBSs are deployed into one macrocell randomly, where the number of SBSs is greater than or equal to the one of IMDs. At the same time, we assume that ${\alpha}_{k}=10$ for any IMD $k$. In addition, ${{n}_{1}}=0.8$, ${{n}_{2}}=0.8$, ${{n}_{3}}=0.3$, ${{n}_{4}}=0.3$, ${{n}_{5}}=0.6$, ${{n}_{6}}=0.03$, ${{n}_{7}}={{10}^{-5}}$, ${{y}_{1}}=0.01$ and ${{y}_{2}}=0.25$ \cite{MYLi2004Sep}; $\kappa^\text{max}=0.9$, $\kappa^\text{min}=0.4$, ${w}_{1}=2$ and ${w}_{2}=2$; $\bar{\varepsilon }=10^{-25}$, ${\hat{\varepsilon }_{j}}=1$ W/GHz for BS $s\in \mathcal{S}$ \cite{YDai2018Dec,YTen2019Jun}; ${{\mathfrak{u}}_{1}}=15$ and ${{\mathfrak{u}}_{2}}=5$ \cite{FVanddenbergh2002}. Moreover, the other important parameters are summarized in TABLE \ref{tab1}, where ${{\ell }_{nk}}$ denotes the distance (in km) between BS $n$ and IMD $k$.
\begin{table}[]
\centering
\caption{SIMULATION PARAMETERS}
\begin{tabular}{cc}
\hline
\rule{0pt}{8pt}\textbf{Parameter}  &\textbf{Value} \\
\hline \rule{0pt}{8pt}
\rule{0pt}{8pt} System bandwidth  & 20 MHz \\
\rule{0pt}{8pt} Noise power & ${10}^{-11}$ mW\\
\rule{0pt}{8pt} Deadline ${T_{i}^{\max}}$ of IMD $i$  & 5$\sim$10 s \\
\rule{0pt}{8pt} The size $Z$ of population & 64 \\
\rule{0pt}{8pt} Wired backhauling rate ${r}_{0}$  & 1 Gbps \\
\rule{0pt}{8pt} Computation capacity $F_{j}^{BS}$  of BS $j$  & 20 GHz \\
\rule{0pt}{8pt} Computation capacity $F_{i}^{UE}$ of IMD $i$ & 1 GHz\\
\rule{0pt}{8pt} Wired power consumption $\tilde{\varepsilon }$ per second & 1 mW\\
\rule{0pt}{8pt} ${{d}_{ik}}$ of $ k$-th computation task of IMD $i$  & 200$\sim$500 KB \\
\rule{0pt}{8pt} ${{c}_{ik}}$ used for computing one bit of ${{d}_{ik}}$ & 50$\sim$100 cycles/bit\\
\rule{0pt}{8pt} Pathloss between MBS $j$ and IMD $i$ & $128.1+37.6\log_{10}\left( {{\ell }_{ij}} \right)$\\
\rule{0pt}{8pt} Pathloss between SBS $j$ and IMD $i$ & $140.7+36.7\log_{10}\left( {{\ell }_{ij}} \right)$\\
\rule{0pt}{8pt} Log-normal shadowing fading & Standard deviation of 8 dB\\ \bottomrule[0.5pt]
\end{tabular}
\label{tab1}
\end{table}
\par
In order to highlight the effectiveness of HAS, we introduce the following algorithms for comparison.
\par
\noindent
\textbf{\textit{Computation at Mobile Terminals (CMT) \cite{TZou2021Oct}:}} All computation tasks of IMDs are completed by themselves locally.
\par
\noindent
\textbf{\textit{Computation at MBSs (CM) \cite{YDai2018Dec}:}} All computation tasks of IMDs are offloaded to MBSs, and then executed at these BSs. In CM, the computation capacity is allocated to served IMDs proportionally.
\par
\noindent
\textbf{\textit{Hierarchical GA and PSO (HGP) \cite{FGuo2018Dec}:}} To solve the problem \eqref{eq16} by using HGP, traditional GA is utilized to perform the coarse-grained search firstly, and traditional PSO is employed for fine-grained search subsequently.
\par
In the simulation, we mainly investigate the impacts of IMD density $\rho^{UE}$ and maximal allowed transmission power $\hat{p}$ of IMDs on the offloading performance, where the IMD density refers to the number of IMDs at each macrocell; ${p}_{i}^{\max}=\hat{p}$ for any IMD $i$. In addition, the mentioned offloading performance includes the total (network-wide) energy consumption, support ratio and total energy consumption at BSs, where the support ratio refers to the ratio of IMDs whose time is less than deadline to all IMDs.
\begin{figure}[!t]
\centering
\centerline{\includegraphics[width=4in]{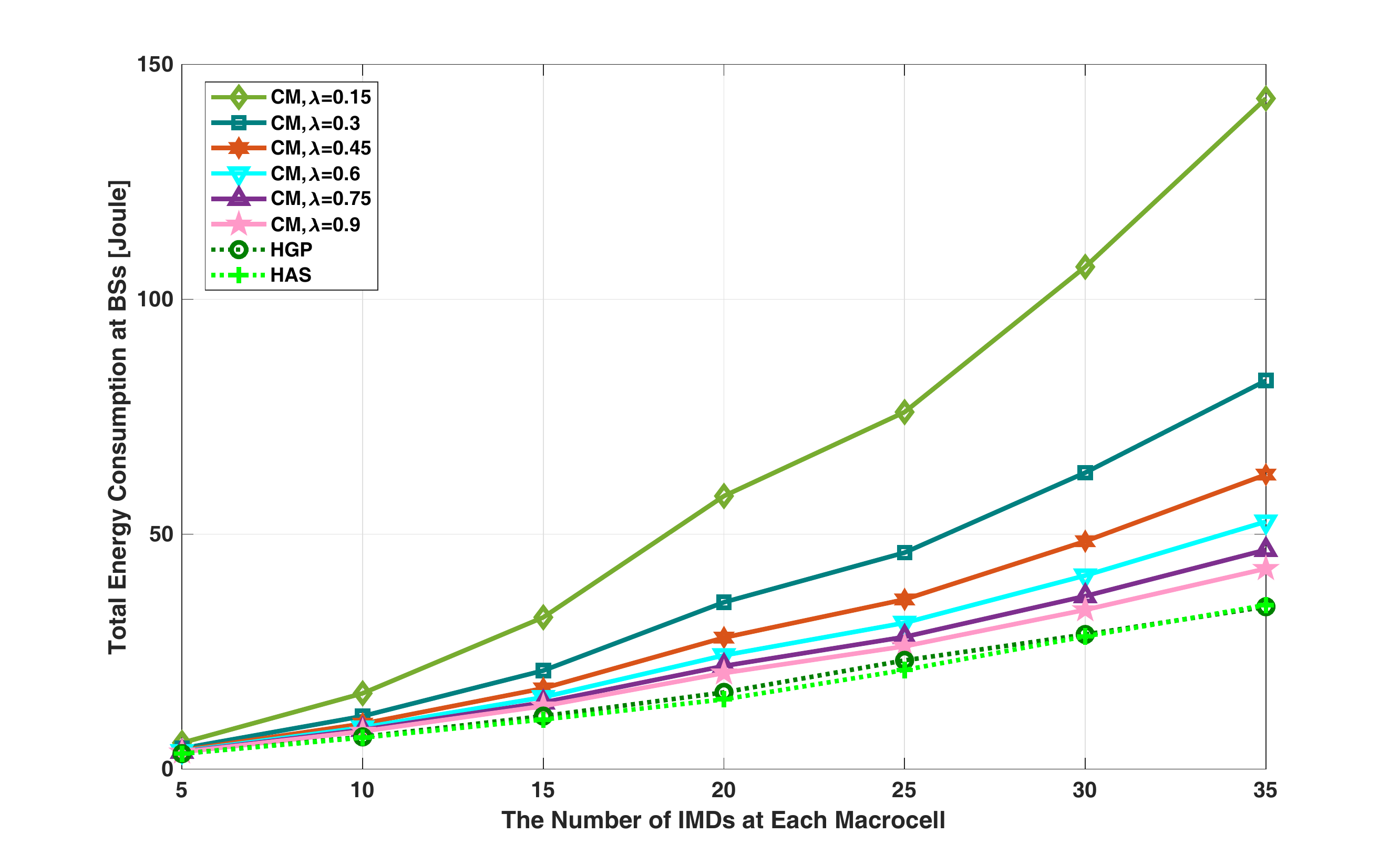}}
\caption{Impacts of IMD density on total energy consumption at BSs under $\rho^{SBS}=35$ and $\hat{p}=23$ dBm.}
\label{fig3}
\end{figure}
\par
Fig. \ref{fig3} shows the impacts of IMD density $\rho^{UE}$ on total energy consumption at BSs under $\rho^{SBS}=35$ and $\hat{p}=23$ dBm, where $\rho^{SBS}$ represents the number of SBSs at each macrocell. As shown in Fig. \ref{fig3}, the total energy consumption at BSs increases with increased IMD density. As we know, when the IMD density increases, more IMDs are served by BSs, and thus total energy consumption at BSs will increase. Seen from Fig. \ref{fig3}, HAS and HGP achieve lower total energy consumption at BSs than CM. That's because HAS and HGP try to minimize the network-wide energy consumption, but CM doesn't concentrate on the optimization of such a performance metric. In addition, HAS and HGP achieve almost the same total energy consumption at BSs. Significantly, in CM, the total energy consumption at BSs increases with decreased frequency band partitioning factor $\lambda$. The reason for this is that a smaller partitioning factor means lower uplink data rates from IMDs to MBSs, and thus the uplink transmission time and energy consumption will increase.
\begin{figure}[!t]
\centering
\centerline{\includegraphics[width=4in]{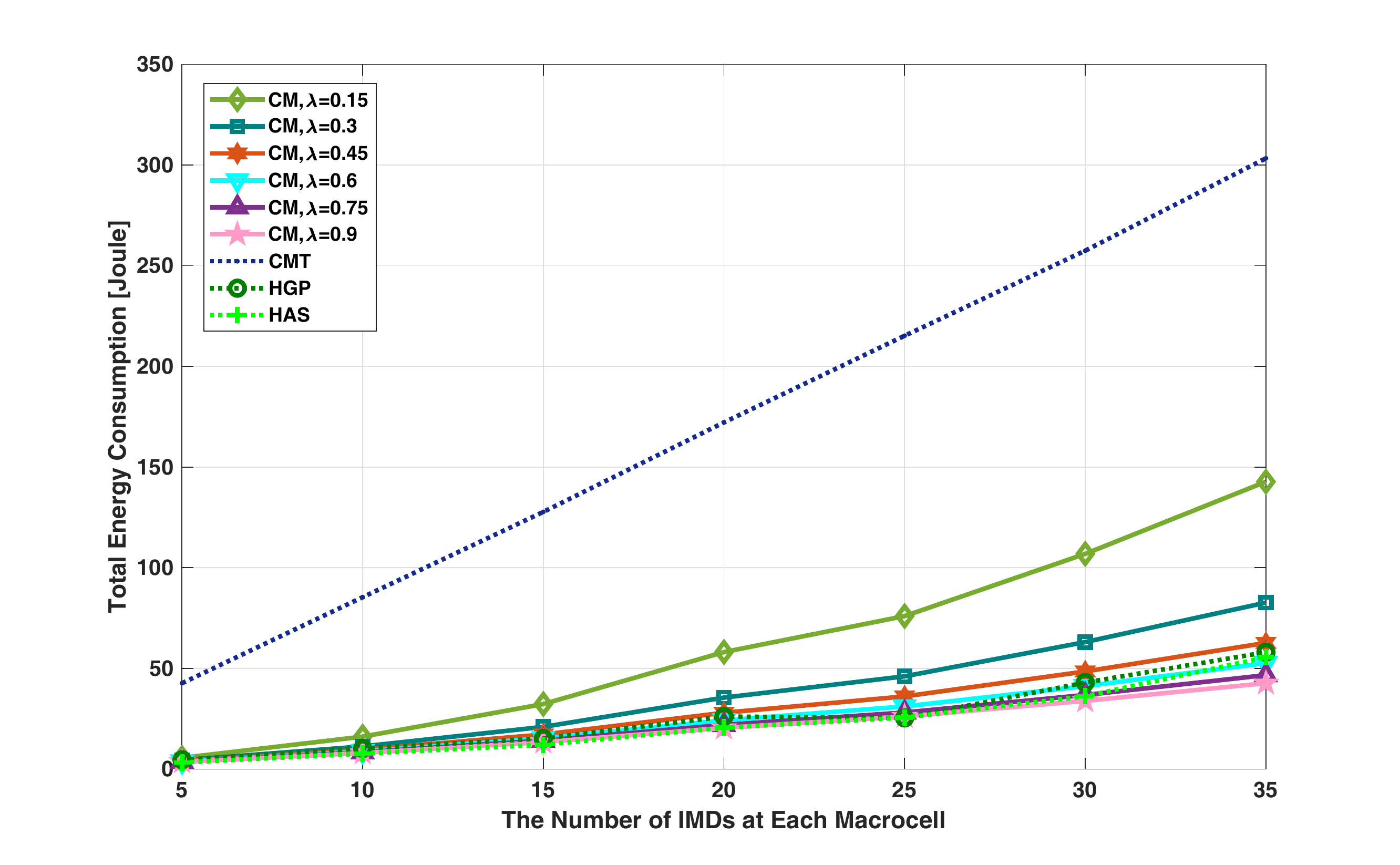}}
\caption{Impacts of IMD density on total energy consumption under $\rho^{SBS}=35$ and $\hat{p}=23$ dBm.}
\label{fig4}
\end{figure}
\par
Fig. \ref{fig4} shows the impacts of IMD density $\rho^{UE}$ on total energy consumption under $\rho^{SBS}=35$ and $\hat{p}=23$ dBm. Similar to Fig. \ref{fig3}, the total energy consumption increases with increased IMD density, and the total energy consumption in CM increases with decreased frequency band partitioning factor $\lambda$. In addition, HAS and HGP almost achieve the lowest total energy consumption among all computation offloading algorithms in the low $\rho^{UE}$ domain. However, they may achieve higher total energy consumption than CMT with $\lambda\le0.75$ since they try to guarantee higher support ratio. Seen from Fig. \ref{fig4}, it is easy to find that HAS may achieve lower total energy consumption than HGP. The reason for this is that HAS avoids the premature convergence and propagate building blocks of best/better individuals.
\begin{figure}[!t]
\centering
\centerline{\includegraphics[width=4in]{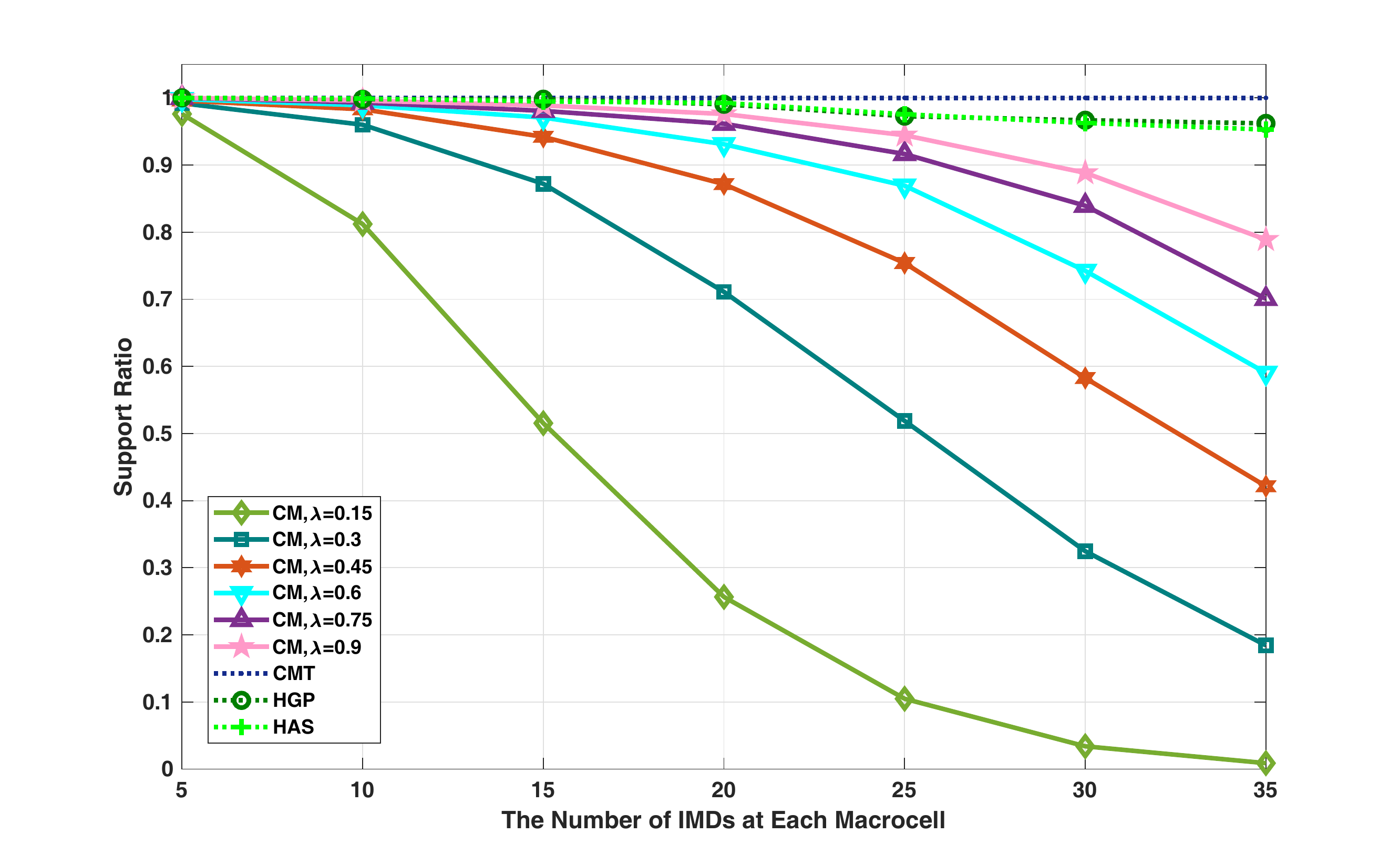}}
\caption{Impacts of IMD density on support ratio under $\rho^{SBS}=35$ and $\hat{p}=23$ dBm.}
\label{fig5}
\end{figure}
\par
Fig. \ref{fig5} shows the impacts of IMD density on support ratio under $\rho^{SBS}=35$ and $\hat{p}=23$ dBm. As shown in Fig. \ref{fig5}, the support ratio decreases with increased IMD density. As we know, more served IMDs often means fewer utilized network resources, and thus the served time increases with increased IMD density. In addition, the smaller frequency band partitioning factor $\lambda$ often means lower uplink data rates from IMDs to MBSs, and thus the uplink transmission time increases with decreased partitioning factor. That is to say, the support ratio of CM increases with decreased factor. Seen from Fig. \ref{fig5}, HAS and HGP achieve almost the same support ratio, higher one than CM, but the lower one than CMT. As we know, CMT has not the uplink transmission and edge computation time, and thus it achieves the highest support ratio among all computation offloading algorithms. Moreover, the support ratio of CMT doesn't change with IMD density since the latter bears no relation to the former.
\begin{figure}[!t]
\centering
\centerline{\includegraphics[width=4in]{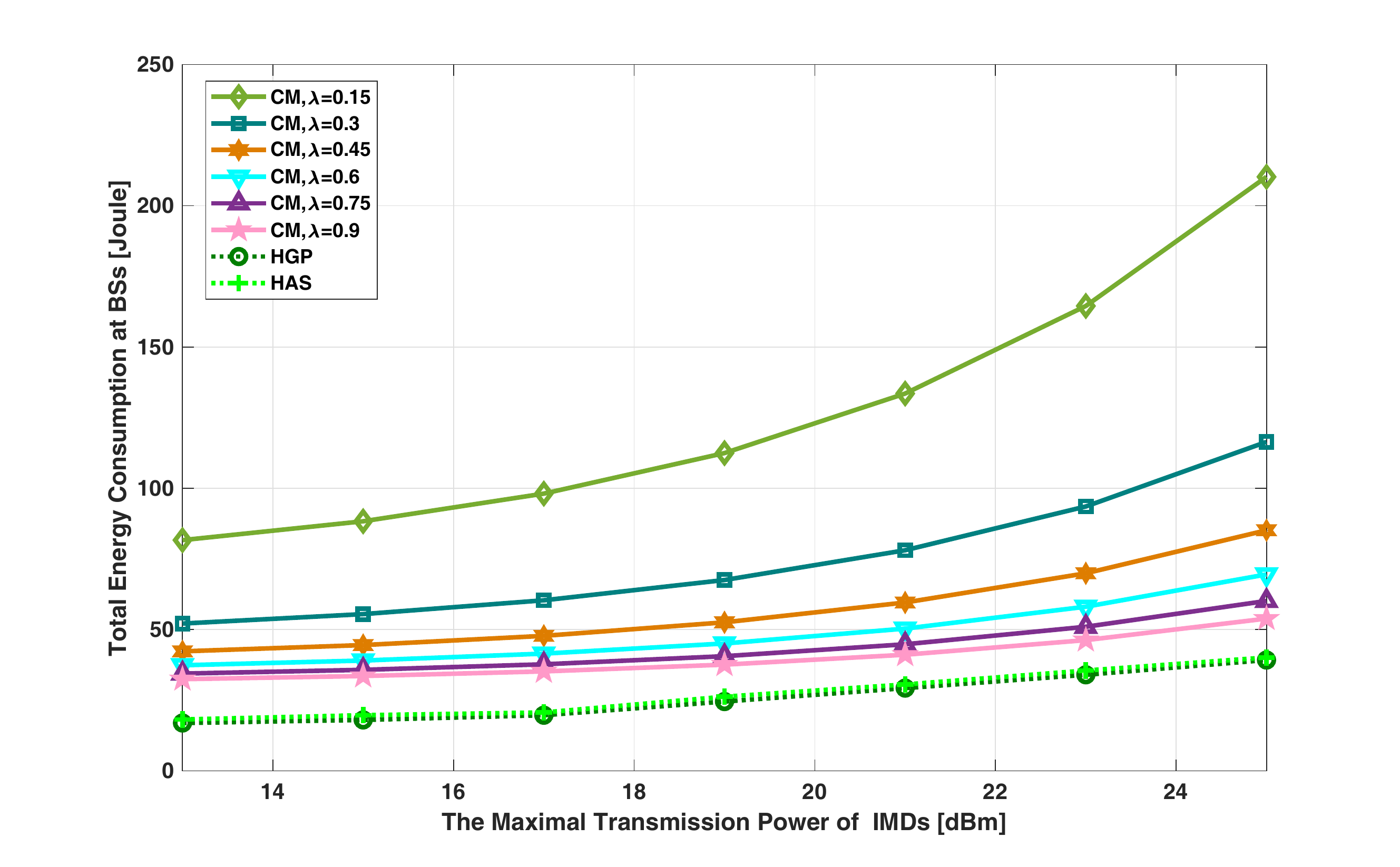}}
\caption{Impacts of maximal allowed transmission power of IMDs on total energy consumption at BSs under $\rho^{SBS}=35$ and $\rho^{UE}=35$.}
\label{fig6}
\end{figure}
\par
Fig. \ref{fig6} shows the impacts of maximal allowed transmission power $\hat{p}$ of IMDs on total energy consumption at BSs under $\rho^{SBS}=35$ and $\rho^{UE}=35$. As illustrated in Fig. \ref{fig6}, the total energy consumption at BSs may increase with increased $\hat{p}$. As we know, the increased $\hat{p}$ results in increased uplink transmission energy consumption, which is contained in the total energy consumption at BSs. Similar to Fig. \ref{fig3}, HGP and HAS achieve almost the same total energy consumption at BSs, and have lower total energy consumption at BSs than CM. In addition, the total energy consumption at BSs increases with decreased partitioning factor $\lambda$.
\begin{figure}[!t]
\centering
\centerline{\includegraphics[width=4in]{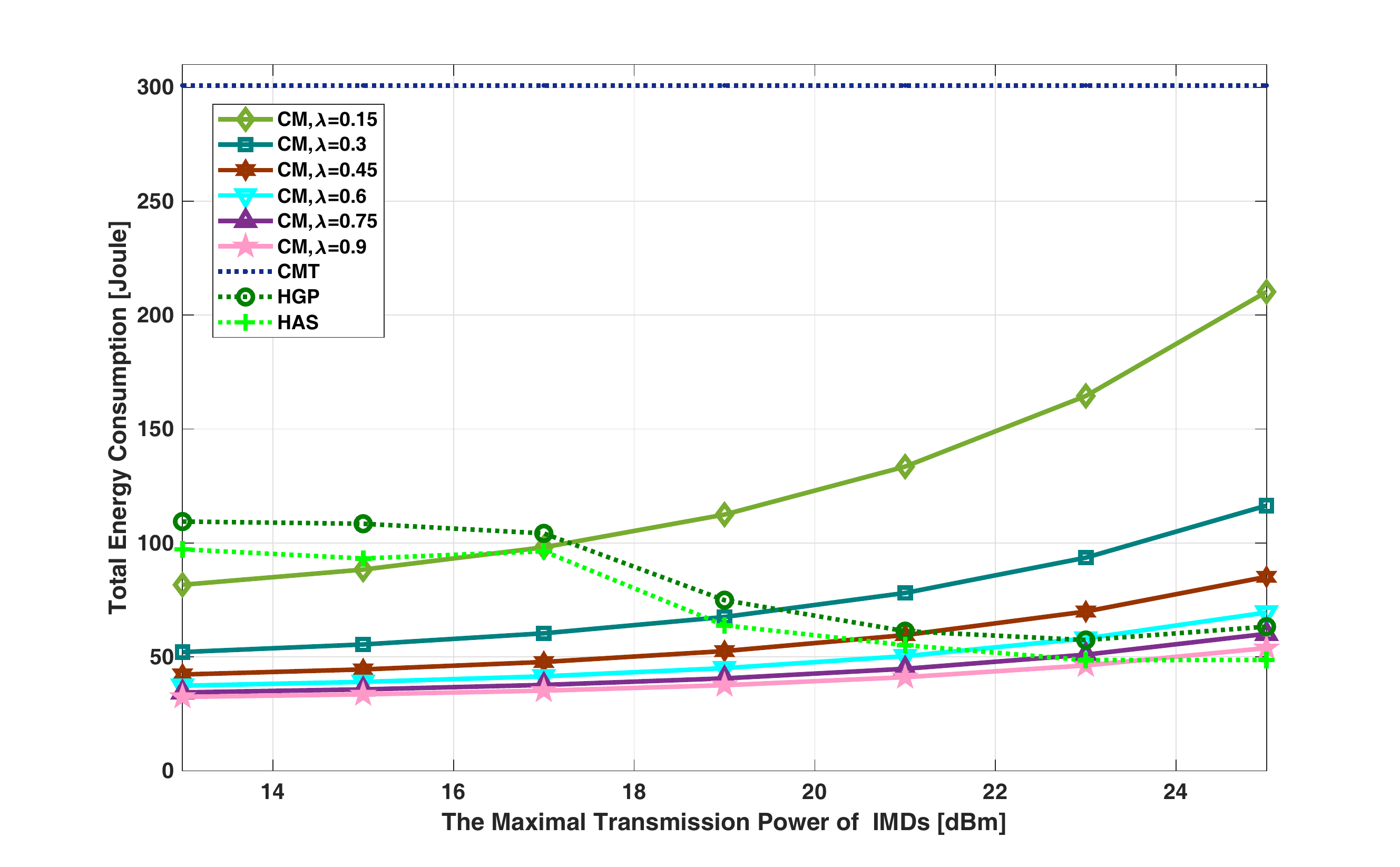}}
\caption{Impacts of maximal allowed transmission power of IMDs on total energy consumption under $\rho^{SBS}=35$ and $\rho^{UE}=35$.}
\label{fig7}
\end{figure}
\par
Fig. \ref{fig7} shows the impacts of maximal allowed transmission power of IMDs on total energy consumption under $\rho^{SBS}=35$ and $\rho^{UE}=35$. As revealed in Fig. \ref{fig6}, the increased $\hat{p}$ results in increased uplink transmission energy consumption, and thus the total energy consumption of CM increases with increased $\hat{p}$. In addition, total energy consumption of CMT should not change with $\hat{p}$ since it has no relation with uplink transmission energy consumption. Seen from Fig. \ref{fig7}, the total energy consumption of HAS and HGP may decrease with increased $\hat{p}$. As we know, when $\hat{p}$ increases, more and more IMDs are served by BSs, less and less IMDs execute their applications by themselves. In addition, in the simulation, we can easily find that the local execution energy consumption of any application is much greater than its energy consumption at BSs. That is to say, the increased $\hat{p}$ results in decreased local execution energy consumption, and finally results in decreased total energy consumption. Similar to Fig. \ref{fig4}, since HAS can avoid the premature convergence and propagate building blocks of best/better individuals, it may achieve lower total energy consumption than HGP. Moreover, the total energy consumption increases with decreased partitioning factor $\lambda$.

\begin{figure}[!t]
	\centering
	\centerline{\includegraphics[width=4in]{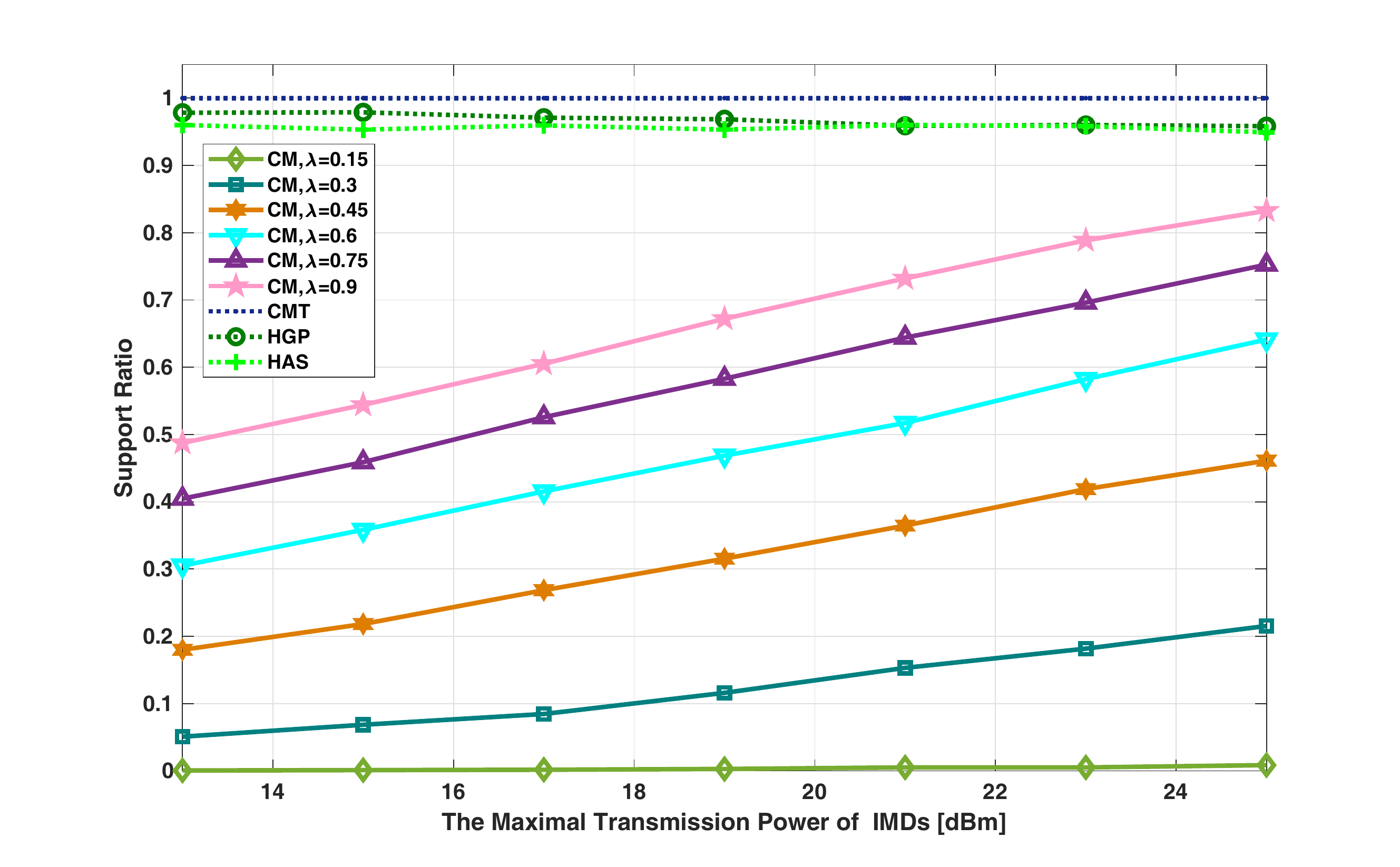}}
	\caption{Impacts of maximal allowed transmission power of IMDs on support ratio under $\rho^{SBS}=35$ and $\rho^{UE}=35$.}
	\label{fig8}
\end{figure}
\par
Fig. \ref{fig8} shows the impacts of maximal allowed transmission power of IMDs on support ratio under $\rho^{SBS}=35$ and $\rho^{UE}=35$. As shown in Fig. \ref{fig8}, the support ratio of CM increases with increased $\hat{p}$. That's because the increased $\hat{p}$ results in increased uplink transmission rate and decreased uplink transmission time in CM. Since CMT bears no relation to $\hat{p}$, its support ratio should not change with such a factor. Seen from Fig. \ref{fig8}, the support ratio of HGP and HAS may slightly decrease with increased $\hat{p}$ since such increased power lets less and less IMDs execute their applications locally. In addition, similar to Fig. \ref{fig5}, CMT achieves the highest support ratio among all computation offloading algorithms, HGP achieves slightly higher one than HAS, and HAS achieves higher one than CM. In addition, the support ratio of CM increases with increased partitioning factor $\lambda$.
\begin{figure}[!t]
	\centering
	\centerline{\includegraphics[width=4in]{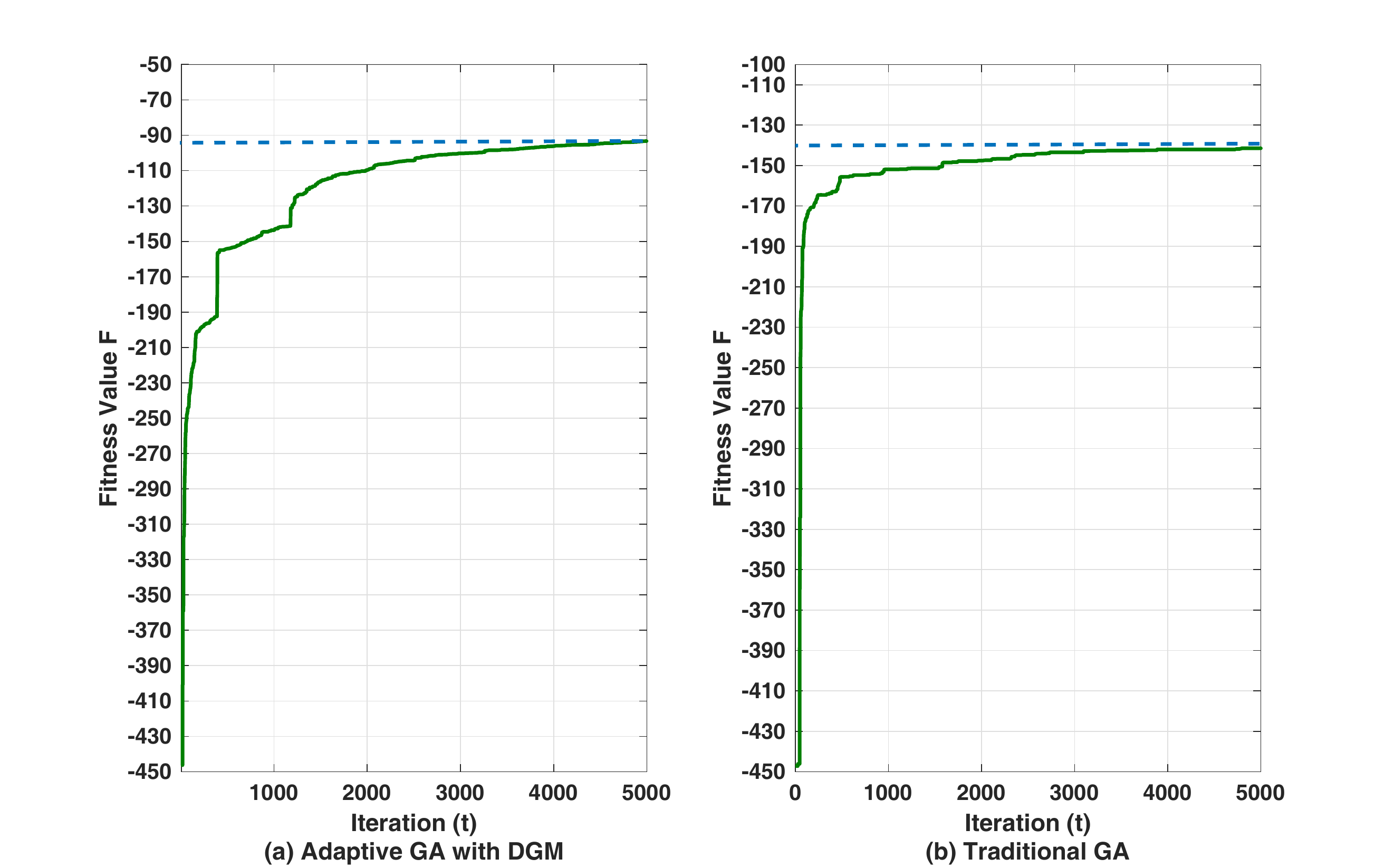}}
	\caption{The convergence of traditional GA and adaptive GA wih DGM.}
	\label{fig9}
\end{figure}
\begin{figure}[!t]
\centering
\centerline{\includegraphics[width=4in]{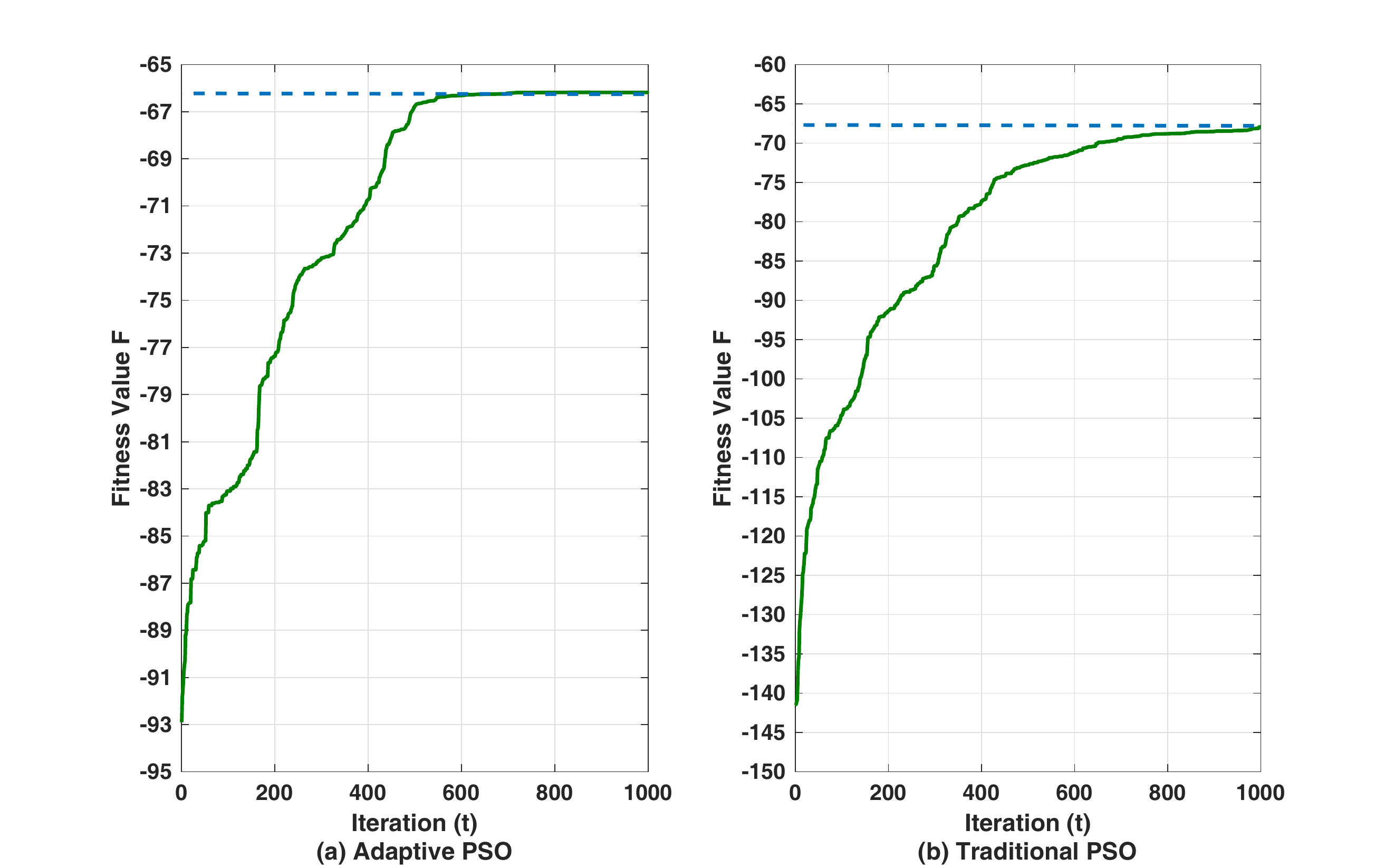}}
\caption{The convergence of traditional PSO and adaptive PSO.}
\label{fig10}
\end{figure}
\par
Fig. \ref{fig9} (a) shows the convergence of adaptive GA wih GDM, and Fig. \ref{fig9} (b) shows the one of traditional GA. Seen from Fig. \ref{fig9}, we can easily find that adaptive GA wih GDM (Algorithm 1) can achieve a better solution than traditional GA. The reason for this is that adaptive GA wih GDM can avoid the premature convergence and propagate building blocks of best/better individuals.
\par
Fig. \ref{fig10} (a) shows the convergence of adaptive PSO, and Fig. \ref{fig10} (b) shows the one of traditional PSO. Seen from Fig. \ref{fig10}, it is easy to find that adaptive PSO (Algorithm 2) can achieve a better solution than traditional PSO. The reason for this is that adaptive PSO can avoid the premature convergence and guarantee local optimum.
\section{Conclusion}\label{sec 7}
In this paper, under the proportional computation capacity allocation and IMDs' latency constraints, the device association, multi-step computation offloading, power control and frequency band partitioning is jointly performed to minimize the network-wide energy consumption for ultra-dense multi-device and multi-task IoT Networks. As for the formulated problem in a nonlinear and mixed-integer form, HAS algorithm is utilized to find its solution. After that the convergence, computation complexity, and parallel implementation analyses are made for such an algorithm. Simulation results show that HAS algorithm can reduce more network-wide energy than other existing algorithms. The future work can include some relevant investigations in NOMA system and/or massive multiple-input multiple-output (MIMO) system.

\end{document}